# Network and Agent Dynamics with Evolving Protection against Systemic Risk

Chulwook Park[1]

[1] International Institute for Applied Systems Analysis (IIASA), A-2361 Laxenburg, Austria.
Correspondence should be addressed to Chulwook Park; parkc@iiasa.ac.at

# Table of contents

**ZVR 524808900**

This research was funded by IIASA and its National Member Organizations in Africa, the Americas, Asia, and Europe.

**UPDATE OR DELETE** Research funded by an external third party has to include the full name of the funder, the title of the project and the grant number, e.g. "The research project 'Forecasting Societies Adaptive Capacities to Climate Change' (FUTURESOC, FP7 230195) was funded by the European Research

## About the authors

**Chulwook Park** is senior researcher at the Seoul National University and was working at IIASA as an IIASA-NRF Postdoctoral Research Scholar. (Contact: parkc@iiasa.ac.at, pcw8531@gmail.com)

## Acknowledgments

This research was supported under the framework of International Cooperation Program managed by National Research Foundation of Korea (Grant Number: 2016K2A9A1A02952017). This research has benefitted from interacting with Ulf Dieckmann, Senior research scholar at the Internal Institute for Applied Systems Analysis; Keivan Aghababaei Samani, Associate Professor at the Department of Physics of the Isfahan University of Technology in Iran as well as with Sara Veradi, who is working in the same department.

# Network and Agent Dynamics with Evolving Protection against Systemic Risk

## Abstract

The dynamics of protection processes has been a fundamental challenge in systemic risk analysis. The conceptual principle and methodological techniques behind the mechanisms involved [in such dynamics] have been harder to grasp than researchers understood them to be. In this paper, we show how to construct a large variety of behaviors by applying a simple algorithm to networked agents, which could, conceivably, offer a straightforward way out of the complexity. The model starts with the probability that systemic risk spreads. Even in a very random social structure, the propagation of risk is guaranteed by an arbitrary network property of a set of elements. Despite intensive systemic risk, the potential of the absence of failure could also be driven when there has been a strong investment in protection through a heuristically evolved protection level. It is very interesting to discover that many applications are still seeking the mechanisms through which networked individuals build many of these protection process or mechanisms based on fitness due to evolutionary drift. Our implementation still needs to be polished against what happens in the real world, but in general, the approach could be useful for researchers and those who need to use protection dynamics to guard against systemic risk under intrinsic randomness in artificial circumstances.

## Introduction

Contemporary social elements are connected through a system of formal and informal flows of risks driven by complex interactions among their structures. It became clear, when seeking fundamental drivers and developing predictive models in order to capture the evolution of systemic risk, that research has moved to a new level [1]. However, as research for this paper expanded into different areas in our attempts to identify universal and domain specific patterns of systemic risk, profound implications for our understanding of real-world dynamic behavior, ranging from protection processes to risk diffusion, became increasingly apparent [2]. Basically, the classical social network metaphor places individuals at the nodes of a network, with the network links representing interactions or connections between those individuals. Networks in any social system, however, are dynamical entities, and in this sense, the practical information of networks is continuously evolving [3]. Here, we developed a model of protection processes against risk diffusion, which along with its dynamics, is shown below.

**Systemic risk across a network:** Systemic risk is a property of systems of interconnected components and can be described as system instability, caused or exacerbated by idiosyncratic events, resulting in potential catastrophe. Notably, in various studies, systemic risk has been blamed for high profile disaster (e.g., for making a significant contribution to the financial crisis of 2008) and also for being a likely cause of cascading failures [4]. A distinguishing feature of such risks, sometimes called network risk, is that they emerge from the complex interactions among individual elements in a system or from their association with each other [3]. The context-varying mechanical flux on a system's risk is actually very complex [5]. The possibility to quantify systemic risk and capture its size need to be performd in the face of all these distortions and patterns of the influences. By knowing the particular form in which an event could trigger instability or collapse an entire system regardless of individual capability at that point, it is possible to quantify with specificity the mechanisms underlying systemic risk using a computerized model. The next step is a simulation, highlighting how the complexity of such interconnected components underpins real-world systemic phenomena, with implications for individual robustness, the propagation of systemic risk, protection flow, and the collective behavior across networks [6]. In particular, a random connectivity pattern proved to be the key to understanding the structure of the network and how elements communicate with each other [7]. The establishment of a simple mechanism (i.e., two types of agents interacting with other within and across their respective social groups) was proposed [8, 9]. At the same time, it was suggested that a strategic decision process be added to explore the influence of a networked interaction on the agents [10]. Other properties of networks applied, such as the concept of evolutionary dynamics, all of which helped us to characterize and understand the architecture of artificial systems from the network property perspective [11, 12].

**Cognitive bias and heuristics**: Individuals have fundamental limitations in terms of their ability to assess probability and situations [13]. Models based on the underlying assumption suggest that individuals perceive their capabilities as becoming more biased if external uncertainty is greater than expected individual capacity. This perception, when distributed in the population, is not quite static [14]. For example, optimal profitability changes if an option – which is present at a given time – has a probability of disappearing or not reappearing in the next unit of time [15]. If there is a low-variance and a high-variance option, the low-variance option is chosen at low reserves and the high-variance option at high reserves [16]. There are several assumptions as to why individuals do not always perceive risk accurately [17]. Evolutionary heuristics, used for the recognition principle, take the best anchoring and adjustment [18]: Charles Darwin's natural selection, Egon Brunswik's texture of natural environments, Roger Shepard's mind as a mirror, Herbert Simon's pair of scissors… comprehensive explanations of behavior are still most often expressed in terms of existing 'inside' the mind. The patterns of information – to which decision mechanisms may be matched – can arise from a variety of processes [19]. Behavior must be explained by an interaction between a heuristic and it's social, institutional, or physical environment: an intuitive mode in which judgements and decisions are made automatically and rapidly [20]. Many papers outline an approach related to research on individual rules of thumb, which has the advantage of fast decision-making based on little information, and also of avoiding overfitting [21]. Individual rules of thumb from individual provide more explicit examples of adaptation because individuals can be studied in the environments in which they evolved [22].

**Imitation and social learning**: A social tool is essential [23]. Its character is vital for social well-being and resilience as it could be damaged or exploited, thus, formulating such a tool is challenging [24]. An unlimited variety of network dynamics and potentials are robust amplifiers of a cascade. It is therefore necessary to consider how the adaptive heuristics for inference and preference ties into a given interconnected social, institutional, and physical tool to produce adaptive behavior. To adequately assess the process, the progress must be made computationally, for example, by performing agent-based simulations of learning agents with evolving properties applied a coexisting macro-scale structure with individual interconnected at the micro-level in a network. An agent-based model can (i) capture emergent phenomena; (ii) provide a natural description of a pattern of behaviour; and (iii) allow realistic understanding of adaptation by incorporating behavioural algorithms into network dynamics [25]. Its description deals with state per time as the determining factor in its allocation of neighbors scheduled for a given moment (cultural evolution). As the agents represent individuals that occurred from the bottom up, the actual state of their behaviour tends to be more informative [26]. This realistic simulation may allow the effects of agents's different behavioral strategies to be tested and monitored (assessment of strategy and heuristic). As the topology of the interaction trait can lead to significant deviations from the predicted pattern of behaviour, it may generate various effects that mimic the behavior of real individuals, allowing sensible decisions to be made based on studying the interactions that an has in the artificially designed system [27]. Moreover, sophisticated behavioral adaptations by individuals are thought to reflect the power of the cultural evolutionary process, in that successful behaviors are copied by other individuals and then propagate strategically through imitation and social learning, rather than through individuals' inherited traits [28]. We should expect individuals to have evolved a set of learning mechanisms that typically enables them to perform well on across a range of different circumstances [29]. These mechanisms encompass imitation and exploration in responding to current stimuli, subject to sensory biases, and learning rules regarding adjusting their behavior in response to nearby individuals.

**Gap statement:** An individual condition of choices and opportunities is offered to account for the structure of the field and to reinforce their [30]. For example, evolutionary explanations regarding systemic risk show how optimal decision-makers are constrained to a biased estimate of their capability and that individuals do alter their strategy according to the perceived resource value [31]. Standard evolutionary models in complex environments provide potentially different biases in decision-making, exposing different experimental groups at different transition probabilities [32]. The computational modeling technique, however, lacks a bridge between the dynamics of agent nodes (of which the fundamental element is a vertex) and the emergent properties of networks. As most tools for laying out networks are variants of the algorithm, it is hard to use them to explore how conditions of a network affect the network's dynamics [33]. While the assessment process is indeed capable of observing at macro-scale for input performance, approaches to addressing the micro-scale to simultaneously obtain more detailed insight need to be treated within the structure of the network itself [34]. This requires large data repositories to be combined to construct representations of trajectories that can be analyzed from different scales and perspectives. Indeed, the mechanisms and the serial algorithm that underpin our understanding of systemic risk in networked agents is still a work in progress. The facts might lead us to find common ground regarding integration of knowledge and methodologies, agreement on definitions, and reconciliation of approaches that many fields have adopted to study networks, all of which present the difficulties and traps inherent in interdisciplinary work.

**Purpose and value:** To identify (a) biases in the assessment of systemic risk, (b) factors influencing these necessary concepts which are mentioned above. Thus, the primary purpose of this study is to examine mechanisms for the evolutionary origin of bias and make a heuristical assessment of protection against systemic risks in a contagious network. We established a modeling framework that can account for the quantitive measurement in agented networks, which allows us to explore on a macro-and micro-scale how protecting potential affects the risk potential. The mechanism tests what we can clearly state for different values of probability and how protection could be drawn by a set of entities against a cascading failure. To reach a better assessment of the risk and how to reduce it, this model not only enables us to directly observe the spread of failure in agented network industries but also to understand how the evolutionary heuristics protects against that spread.

## Methods

*Network properties.* We consider an Erdös-Rényi network [35] with a given number $n$ of nodes, connection probability $p_c$, and resultant adjacency matrix $A$. Each node can be in one of two states: not failed or failed. All nodes are initially without failure. *Agent properties.* One agent is associated with each node and is characterized by its capital and strategy (see below). *Payoff dynamics.* In each time step, each agent receives one unit of payoff, which is added to its capital $c$, of which fractions $f_m$ and $f_p$ are spent on maintenance and protection, respectively, resulting in the updated capital $1 + (1 - f_m - f_p)c$. *Failure dynamics.* In each time step, a failure potential can originate at each node with probability $p_n$ and can propagate along each link with probability $p_l$. A failure potential turns into a failure with probability $1 - p_p$, depending on an agent's investment into protection; a possible choice is $p_p = p_{p,max}/(1 + c_{p,1/2}/(f_p c))$. A failure lasts for one-time step and causes the loss of an agent's capital. *Strategy dynamics.* Each agent chooses its protection level according to the heuristics $f_p = f_{p0} + f_{p1} C$ truncated to the interval $(0, 1 - f_m)$, where $C$ is a measure of the centrality of the agent's node normalized to the interval $(0,1)$. The strategy values $f_{p0}$ and $f_{p1}$ evolve through social learning and strategy exploration as follows. In each time step, each agent with probability $p_r$ randomly chooses another agent as a role model and imitates that agent's strategy values with probability $p_i = 1/(1 + \exp(-s\Delta c))$, where $s$ is the strength of selection and $\Delta c$ is the difference between the role model's capital and the focal agent's capital. In each time step, each agent with probability $p_e$ randomly chooses one of its two strategy values and alters it by a normally distributed increment with mean $0$ and standard deviation $\sigma_e$ (see Supplement information for more detail).

## Results

To observe the process of dynamics, the model uses an array as a probability of failure with a given number of initially influenced nodes. With the set of features regarding the protection dynamic applied according to the model description (see Methods), the simulation results are organized as follows; Section 1 describes the fundamental characteristics of risk diffusion in a random networked system. Section 2 investigates the framework that allows us to look the assumption in a tractable way while imposing realistic protection against the failure probability through social learning on agents. Section 3 characterize their stationarity from the observations of applied dynamics over time to see how the spread of the failure happens.

**Part1: Fundamental structure (in random network)**

To start with, individuals in the model are considered as vertices (fundamental element drawn as nodes) and a set of two elements drawn as a line connecting two vertices (the lines are called edges) depending on the information in the graph [usually controlled by (*n, p*)]. There are two parameters: the number of nodes (*n*), the probability that an edge is present (*p*).

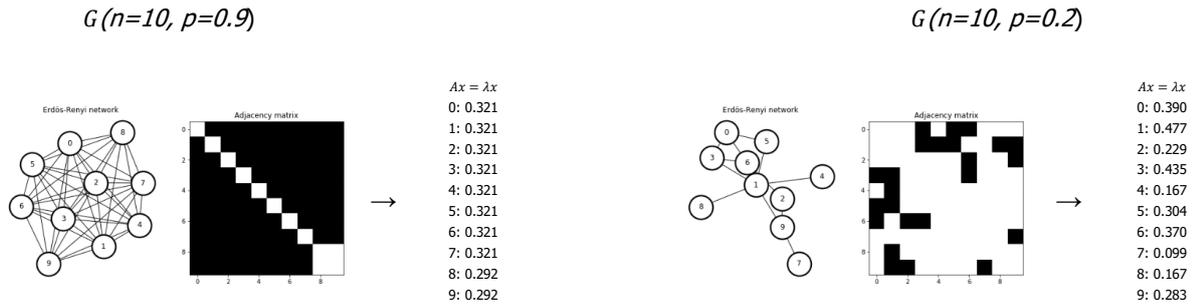

Figure 1.1. A prototype of the random network with its property. Number of nodes $n$ = 10, Connection probability $p$ = 0.9 (left set), $p$ = 0.2 (right set). At each section, the plots of the left side show the random (Erdös-Rényi) network created. A circle represents each node with an arbitrarily assigned label from 0 to 9, and each line represents a link. The plots of the middle show their adjacency matrix with its entry in row $m$ column $n$ (either 1=black or 0=white) corresponding to its eigenvector centrality (right side).

The results from Figure 1.1 follow from a standard representation in graph theory. It takes into account the fact that a higher-degree node has a higher chance of being connected to an agent in a simple way through the degree distribution. For a network artificially produced as $G(n,p)$, $n = 10$ and $p = 0.9$ as an example; The number of possible edges is in $n(n-1)/2 = 10*9/2 = 45$. The expected average of node degree is $p(n-1) = 0.9*9 = 8.1$. To quantify the probability that a node has degree $d$ for all $[0 \leq d \leq (n-1)]$ note that a node has degree zero if nothing is connected to it. A node has a degree $(n-1)$ if all nodes are connected to it. For a node to have degree $d$ in a network with $n$ nodes, there must be $d$ 'connections' and $(n-1-d)$ 'nodes that are not connected to it.' Since the probability of a 'connect' is $p$, the probability of a 'not connecting to a node' is $(1-p)$. The outcome $d$ 'heads' and $(n-1-d)$ 'tails' occurs with probability $p^k(1-p)^{n-1-d}$, but there are '$(n-1)$ choose $d$' ways in which this outcome can occur (the order of the flip results does not matter). The probability that a given node has degree $d$ is given by the binomial distribution $\binom{n-1}{d}p^k(1-p)^{n-1-d}$, simply is $\mu = p(n-1)$ which is also the average node degree. Depending on the degree of each node, the average path length and time between connected nodes can also be calculated. For small values $[p(d)]$, the graph shows a small number of edges, isolated clusters, and a large different between path lengths and path time. For large values $[p(d)]$, the graph shows almost a complete graph, and the path length and time shorten. In network analysis, as indicators of centrality identify the most important vertices (nodes) within a graph, applications include identifying the most influential node(s) in a network. The eigenvector centrality for node is $(Ax = \lambda x)$, where $A$ is the adjacency matrix of the network with eigenvalue $\lambda$. The principal has an entry for each of the $n$-vertices. The larger the entry for a vertex is, the higher the ranking with respect to eigenvector-centrality. With respect to the fundamental characteristic of the model, we implement that an individual (node) can catch a failure if one of its neighbors is infected to measure how cascades of failure can propagate through the network.

**Proposition of the cascading failure**: The model uses an array (vector) as a probability of failure $[p \in (0,1)]$ with the initially given influenced nodes ($1 \leq j \leq N$) being noted merely by ($p\_j$). Each node can be in one of two states; not failed or failed. All nodes are initially without failure. The fundamental characteristics of network (determining the failure $S_{ji}$) are obtained from each link of nodes in the current context. The elementary level of risk depends on the network units, which depend on the co-occurrence of the $i$ and $j$ of the nodes. This reflects that individuals are more biased when an individual is highly linked in its network. In what follows, such a probability of failure will be determined by the number of links from the node of the specification scaled by $R/S$. If we keep the individual characteristics as constant ($K$). $R/S$ is equal to the risk [failure probability: $p \in (0,1)$] as a function of connectivity which was created by the eigenvector centrality for nodes ($\lambda x$), and $R$ is just equal to $K$ over $S$ ($R = K/S$). Nodes at lower (higher) links should have a lower (higher) connectivity with their risk, and vice versa. In other words, if we remove nodes from the network, the bias reduces where the links have decreased, even if they have retained their individual characteristics throughout the entire process. The intuition behind these results is that the higher-degree agents are more exposed to cascading failure risk than the lower-degree agents; this increases potential for cascading failure.

**Part 2: Protection against the failure (imposing realistic dynamics)**

In line with this observation, protection dynamic was applied against the risk of failure. The model allows an agent to make a costly investment into protection. Note that we assume the systemic risk as failure potential from the dynamics turns into a failure with probability $1 - p_p$ depending on an agent's investment into protection; $p_p = p_{p,max}/(1 + c_{p,1/2}/(f_p c))$ when a failure lasts for one time step.

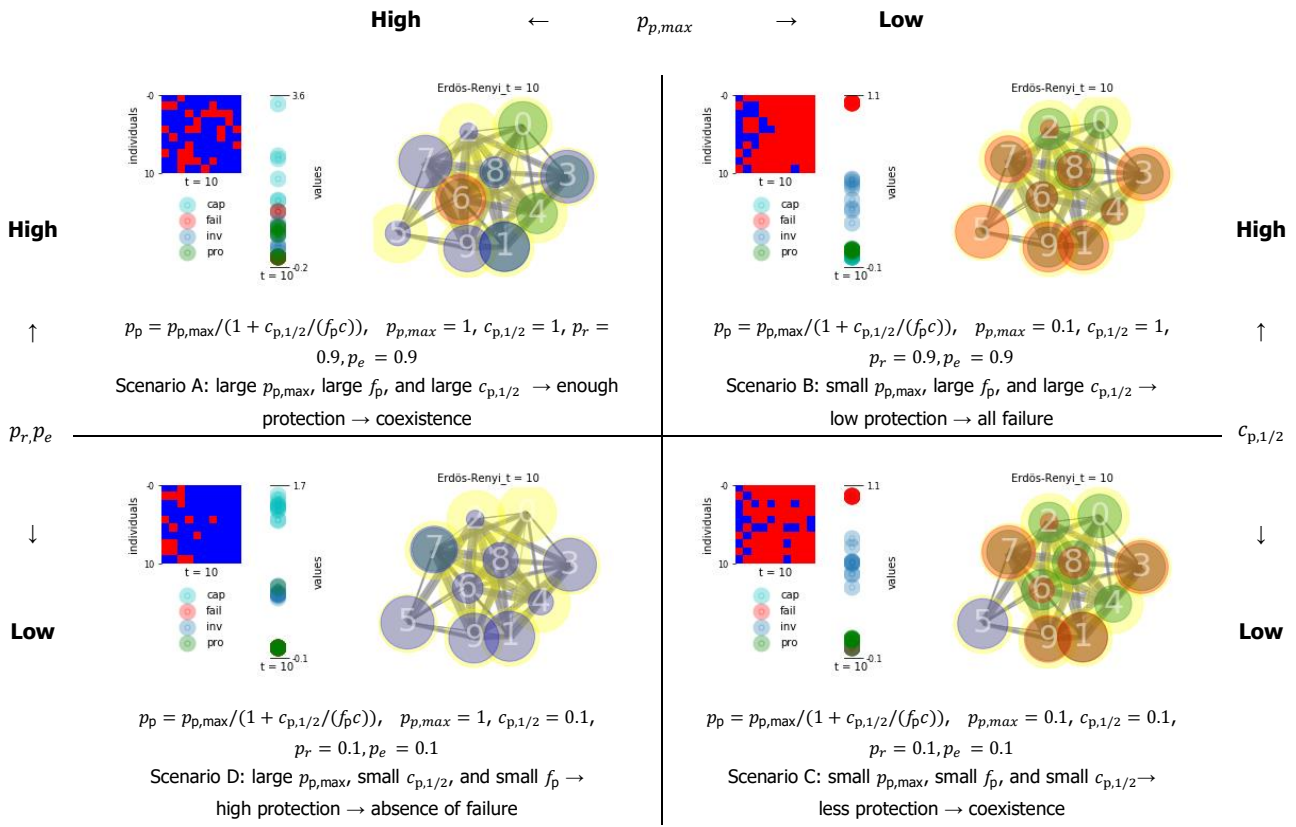

Figure 2.1: Protection dynamics against systemic risk. At each section, plots of the left-hand side show the matrix [horizontal axis = time step from 1 to 10, vertical axis = individuals (10), color of the matrix = failure state between fail (red) and absence of fail (blue)]. The plots of the middle show each individuals' parameter values at the time step [$t$=10: cyan: cap=capital ($c$), red: fail=failure, blue: inv=investment ($f_p$), green: pro=protection potential ($p_p$)]. The plot on the right-hand side represents individuals' dynamics within a random network; node number = random label of each node, line width = eigenvector centrality, node color = states [failure (red) ←→ (gray) absence of failure, green = protection potential, yellow = initial structure of the state without failure and protection]. Initialized parameters of the simulations are: nodes $n$=10, connection $p$=0.9, p=1, $f_{p0}$=0.4, $f_{p1}$=0.5, $f_m$=0.1, $s$=1, $\mu$=0.0, $\sigma$=0.1, $p_n$=0.1, $p_l$=0.3, $t$=10.

***Framework***: As a basic framework of this model, the probabilities of these plausible scenarios can be proved as follows: first, scenario A's failure potential $1 - p_p$ becomes 0.527 because protection $p_p$ is 0.473 based on the possible choice $p_p$ is $p_{p,max}/(1 + c_{p,1/2}/(f_p c)) = 1/(1 + 1/(0.9*1))$ with a given parameter values ($p_{p,max} = 1, c_{p,1/2} = 1, p_r, p_e = 0.9, C = 1$). Second, scenario B's failure potential $1 - p_p$ becomes 0.953 because the protection $p_p$ is 0.047 based on the possible choice $p_p$ is $p_{p,max}/(1 + c_{p,1/2}/(f_p c)) = 0.1/(1 + 1/(0.9*1))$ with a given parameter values ($p_{p,max} = 0.1, c_{p,1/2} = 1, p_r, p_e = 0.9, C = 1$). As can be seen on the right-hand-side upper plot, this scenario obtains the worst result corresponding to all failure as time goes by. Third, the scenario C's failure potential $1 - p_p$ becomes 0.95 because the protection $p_p$ is 0.05 based on the possible choice $p_p$ is $p_{p,max}/(1 + c_{p,1/2}/(f_p c)) = 0.1/(1 + 0.1/(0.1*1))$ with given parameter values ($p_{p,max} = 0.1, c_{p,1/2} = 0.1, p_r, p_e = 0.1, C = 1$). Finally, scenario D's failure potential $1 - p_p$ becomes 0.5 because the protection $p_p$ is 0.5 based on the possible choice $p_p$ is $p_{p,max}/(1 + c_{p,1/2}/(f_p c)) = 1/(1 + 0.1/(0.1*1))$ with a given parameter values ($p_{p,max} = 1$, $c_{p,1/2} = 0.1,$ , $p_r, p_e = 0.1, C = 1$). As can be seen on the left-hand side of the bottom plot, this scenario obtains the best result corresponding to the absence of failure. Moreover, even in the observed coexistence scenario A ($p_{p,max} = 1, c_{p,1/2} = 1, p_r, p_e = 0.9, C = 1$), our model shown in Figure 2.2 reflects that the contagion or systemic risk is also likely to be intensified with recovery time, causing a significant costs of the failures. We have observed that immediate or procrastinate of intervention provide the associated with the propagation criteria in micro-scale for each node.

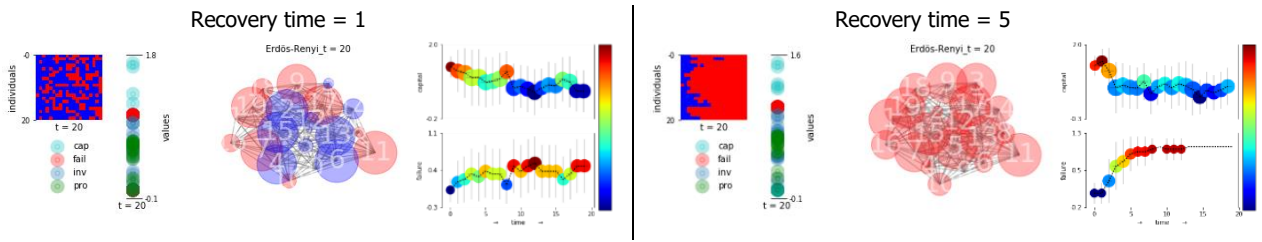

Figure 2.2: Protection dynamics against systemic risk with the recovery time delay. With the same initialized parameter values (Scenario A: coexistence), the set of the left-hand-side plots denotes no-time-delay (=1) case for the recovery time against the failure potential, while the right-hand set represents the time-delay (=5) case. At each section, plots of the left-hand side show the matrix [horizontal axis = time step from 1 to 20, vertical axis = individuals from 0 to 20, color of the matrix = failure state between fail (red) and absence of fail (blue)] corresponding to each individuals' parameter values at the time step [t=20: cyan: cap=capital ($c$), red: fail=failure, blue: inv=investment ($f_p$), green: pro=protection potential ($p_p$)]. The graph on the middle represents their dynamics with a random network; node number = random label of each node, node color = states (failure = red, absence of failure = blue). The plots on the right-hand side show the capital (upper) and failure (lower) trend according to time steps; marker = averaged value according to the time steps (1~20), marker size = variability, dotted line and color of the marker = strength (color bar) of the averaged value.

***Recovery rate***: A sharp differentiation is possible among individuals. Obviously, such nodes that have immediate recovery (left-hand-side plots) seems to have potential to protect against the propagation of failure; on the other hand, nodes that have a malfunction (right-hand-side plots) seem to not have enough potential. Statistically, the simulation consists of repeated trials, with each trial having two possible outcomes. One of the consequences can be called failure (=red) and another can be called absence of failure (=blue). A probability of failure is the same for every trial, like the flipping of coins $n$ number of times, as it is based on the binomial variable which we defined as this model's basic structure. The probability of failures in each trial as given by $[P(X) = \frac{n!}{r!(n-r)!} p^r (1-p)^{n-r} =$ $C(n,r)p^r(1-p)^{n-r}$]. Where, $n$ is the total number of trials, $r$ is total number of failure events, and $p$ is the probability of failure on a single trial. In the plots of the left-hand side in Figure 2.2, we suppose the probability of absence of failure to be about zero point six and the probability of failure to be zero point four [from the Scenario A case probability is calculated by the number of failures from 20 time steps: $f_j = \frac{h_j}{N}$, $h_j = \sum_{k=1,\dots,N;X^{(k)}=j} 1$].

We assumed a random variable $X$ as being equal to the number of failures after 20-time steps. (i) One of the first conditions of the result is that it is made up of a finite number of independent trials. This means that the probability of whether we obtain failure or absence of the failure on each trial is independent of whether we simply obtained failure or the absence of the failure on a previous trial. Thus, in the case of the left-hand-side plots with recovery in every time steps (immediate intervention), the simulation results are made up of independent trials. (ii) Another condition is that each trial clearly has one of the two discrete outcomes in which the variable $X$ should be clearly classified as either a failure or an absence of the failure with (iii) a given fixed number of trials. Then (iv) the final condition of the probability of failure (=0.4) and absence of the failure (=0.6) on each trail is constant, which we have already measured on each trial from the Scenario A case. In the plots on the right-hand-side case in Figure 2.2, however, the probability would no longer be the same but would change from trial to trial. We have the variable $X$ which is equal to the number of failures from a designate population. This looks like as if it could be the same operation because, in it, each trial can be classified as either a failure or an absence of the failure over a fixed number of trials (=20). At the same time, there is a probability of the variable $X$ not being constant at each trial because of the recovery delay which is not made up of independent trials. The probability of the failure or the absence of the failure on the first trial would be equal to the whole number of individuals between the two simulation cases [$P(k \text{ on } 1^{st} \text{ trial}) = 0.4$], but the probability of the second trial (and the following one) would be not the same since the simulation of the lower case but depends on what happened on the first trial [$P(k \text{ on } 2^{nd} \text{ trial}) \neq constant$]. Simply put, each trial is being carried out without replacement, and this causes an exponentially large difference between two cases. In other words, this does not meet the independent trial condition, and the probability in the next trial is dependent on what happened on the previous trial. As replacement is not taking place, the probability of failure on each trail also is not constant, in contrast to the probability of the failure being constant on every trial in the case of the upper simulation [$P(k \text{ on } 2^{nd} \text{ trial}) = constant$]. Thus, to reduce the potential ramifications from such additional losses to the others, an intermediate intervention in terms of recovery may be preferred to the potential damage from individual failures and may also guarantee that it is strategically possible to recover even large insolvent individuals with losses to uninsured connectors. Before the potential for failure can advance the value of their propagation, they must be identified, and the recovery value of the individual capital estimated.

**Protection level**: Inspired by the plausible scenarios presented above which included recovery delay, we developed our focus on the parameter of ($f_p c$) with a given of more individuals and time steps (nodes=100, $t$=100) without recovery delay. As will be noted, this is because of the applied function of $p_p$ [$= p_{p,max}/(1 + c_{p,1/2}/(f_p c))$] on this application will be decided by [$p_r, p_e, C \rightarrow (f_p c)$] when we consider the $p_{p,max}$, $c_{p,1/2}$ and time delay (=1) as constants. This refers to an investment made by an agent (we define nodes as agents, as agents make decisions regarding investment in protection) in order to protect itself against the risk of failure in terms of how the failure propagation mechanism influence the agents' decisions as they generate.

Figure 3.1 shows different evolutionary observations parameterized $p_{p,max} = 1$, $c_{p,1/2} = 0.5$ as constant and changes only the protection values controlling imitation and exploration probabilities ($p_r$, $p_e$) with the eigenvector centrality ($C$). A possible scenario of the different conditions is as follows: scenario A = strong connection with strong imitation and exploration [$p_p = p_{p,max}/(1 + c_{p,1/2}/(f_p c))$, $p_{p,max} = 1$, $c_{p,1/2} = 0.5$, for $(f_p c)$ $p = 0.9, p_r = 0.9, p_e = 0.9$], scenario B = weak connection but strong imitation and exploration [$p_p = p_{p,max}/(1 + c_{p,1/2}/(f_p c))$, $p_{p,max} = 0.1$, $c_{p,1/2} = 0.5$, for $(f_p c)$ $p = 0.1, p_r = 0.9, p_e = 0.9$], scenario C = weak connection with weak imitation and exploration [$p_p = p_{p,max}/(1 + c_{p,1/2}/(f_p c))$, $p_{p,max} = 1$, $c_{p,1/2} = 0.5$, for $(f_p c)$ $p = 0.1, p_r = 0.1, p_e = 0.1$], scenario D = strong connection but weak imitation and exploration [$p_p = p_{p,max}/(1 + c_{p,1/2}/(f_p c))$, $p_{p,max} = 1$, $c_{p,1/2} = 0.5$, for $(f_p c)$ $p = 0.9, p_r = 0.1, p_e = 0.1$].

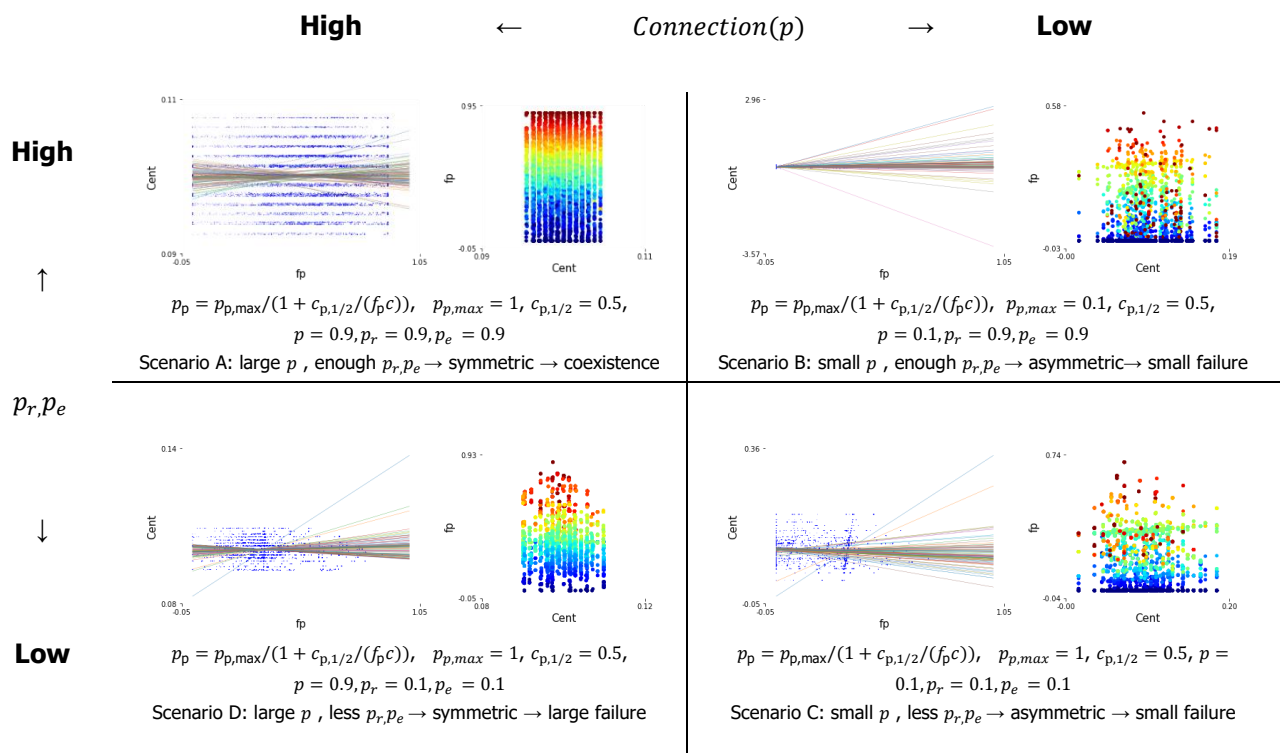

Figure 3.1: Protection dynamics against systemic risk with pattern recognition. At each section, the plot of the left-hand side shows the relation between protection level ($f_p$) and the eigenvector centrality which is controlled by the initialized random network property [$G(n, p) \rightarrow$ Cent = eigenvector centrality] according to time steps (100). The blue dots represent individuals, and the lines represent regressions at each time steps. The plots of the right-hand side at each section represent that large connection probability caused asymmetric pattern with the correlation between the two parameters. Instead, the small connection probability does not.

The patterns observed from the four parametrizations give us different evolutionary patterns according to the time series. For small links (connection $p = 0.1$) among the individuals, a weakly interacting pattern was observed in which the centralities were widely distributed, with strong interactions rather than large links (connection $p = 0.9$) being observed, in which the centralities were tightly distributed. We also discovered that once the link thickness was driven, a strongly interacting regime emerges in the sense that the large connection probability caused a symmetric pattern between the centrality and the protection level, while the small connection probability did not. The results give us some insights into what possible patterns are going to be, with the artificially designated parameters being a plausible concept of the protection factors against the systemic risk.

**Strategies**: Following the previous observation, we now present another analytical characterization to see if an alternative strategy could affect the failure trend based on which strategy values $f_{p0}$ and $f_{p1}$ evolve through social learning and exploration. We start this simulation with the assumption that the environment has a high protection ($p_{p,max} = 1$), strong centrality ($p = 0.9$), and imitation ($p_r = 0.9$) but that an agent does not have enough opportunity to explore the other strategy values ($p_e = 0.1$). As the exploration probability can cause the different protection level in $f_p$, the parameter set assumes that a smaller value of $p_e$, turns out to be incapable of the dynamics expected to result in a protection level. In other words, even if all agents could chooses another agent as a role model and imitates that agents' strategy values (capital) in each time step with the applied function ($p_i = \left[1 + e^{[-s(\pi_r - \pi_f)]}\right]^{-1}$), each agent could rarely alter its strategy value to the other strategy value because there is no chance of exploring the other strategy.

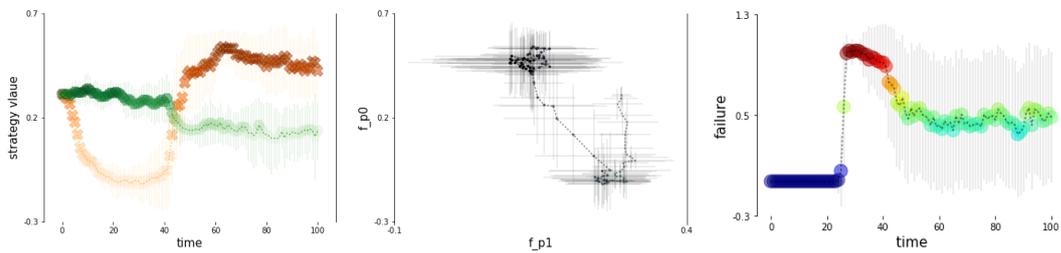

Figure 3.2: Described protection dynamics against systemic risk with small exploration rate. The plot of the left-hand side shows a trend of the two strategies ($f_{p0}$=red, $f_{p1}$=green, dashed vertical lines = variability), corresponding to its evolutionary trajectory (middle side of the plot: dot = averaged value, strength of the dot = time step, dashed horizontal and vertical lines = variability). Note that the plot of the right-hand side represents the failure (vertical axis) and variability (dashed vertical line) at each time step (horizontal axis). Initialized parameters of the simulations are: nodes $n$=100, connection $p$=0.9, capital $c$=1, $f_{p0}$=0.3, $f_{p1}$=0.3, $f_m$=0.1, , $s$=100, $p_r$=0.9, $p_e$=0.1, $\mu$=0.0, $\sigma$=0.1, $p_{p,max}$=1, $c_{p,1/2}$=0.5, $p_n$=0.001, $p_l$=0.1, $t$=100.

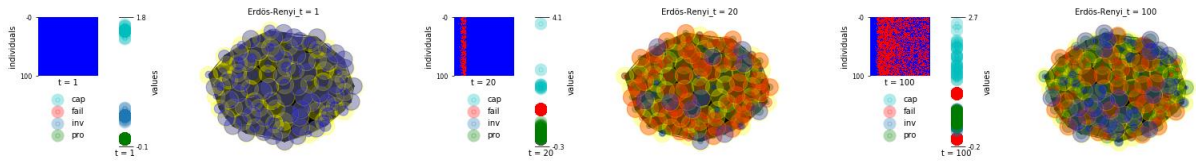

Figure 3.3: Protection dynamics against systemic risk with small exploration rate. At each section, plots of the left-hand side show the matrix [horizontal axis = time step from 1 to 100, vertical axis = individuals from 0 to 100, color of the matrix = failure state between fail (red) and absence of fail (blue)]. The plots of the middle show each individuals' parameter values at the time step [$t$=100: cyan: cap=capital ($c$), red: fail=failure, blue: inv=investment ($f_p$), green: pro=protection potential ($p_p$)]. The plot on the right-hand side represents the individuals' dynamics with a random network; node color = states [failure (red) ⟵⟶ (blue) absence of failure, green = protection potential, yellow = initial structure of the state without failure and protection].

Figure 3.2 shows how different strategies for different evolution in networks with failure. For the most argued dilemmas of social networks, we consider the fraction of $f_{p0}$ as one strategy and the fraction of $f_{p1}$ as the other strategy, averaged over 100 individuals and time steps. We were able to find quite clear correlation between the parameter values (capital and failure, as well as the strategies of $f_{p0}$ and $f_{p1}$). The strategy of $f_{p1}$ multiplied by eigenvector centrality of the $C$ in particular shows different behavior. Although there is a relative score which assigned to the $f_{p1}$ individuals based on the concept that a high eigenvector-centrality score ($C \in$ 0,1) contribute more than the others that have a relatively lower score, this feature does not causes the same trend, and this results in an exponential decay of the trend in applied time series (plot on the left-hand side in Figure 3.2). The simulation results with a small exploration rate reflect that the centrality with the neighbor in the network conceptually matching relations of parameters expand nonlinearly with time, these densifications suggest that the existing structure of the network may constrain what will be happen in the next time step. Evidence shows that biases between parameters occur at very early stages. Hence, the probability of exploration (social learning), which might be another crucial factor in providing resources to violate expectations and lead to novel trends with high impact.

**Part3: Coevoluntionary hueristics (stationary case)**

Regarding the influence of imitation and exploration, we expected to need to observe the evolutionary state in these traits ($f_{p0}$ and $f_{p1}$). We reduced the exploration probability and its normally distributed increment so that the trait variabilities in $f_{p0}$ and $f_{p1}$ become much smaller; we did this by measuring these trait variabilities using trait ranges as a coefficient of variation (standard deviation decided by a mean) of, at most, 10% in either trait.

Divergent　　　　　　　　　　　　　　Convergent

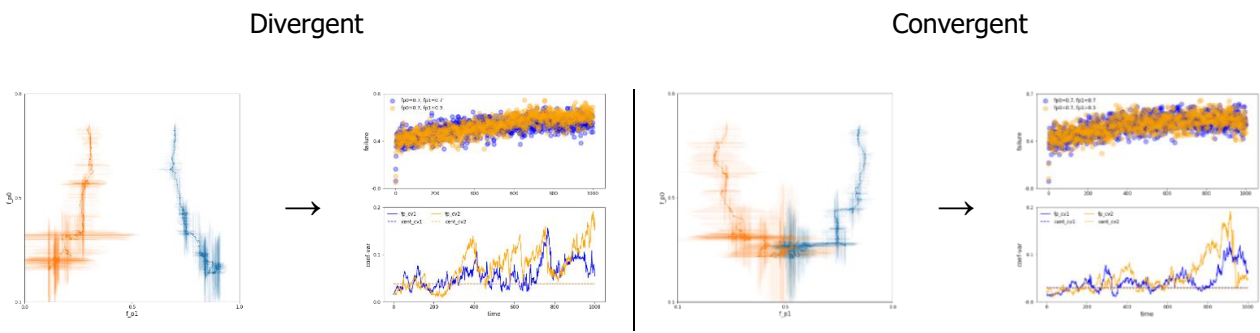

Figure 4: Described protection dynamics against systemic risk with the different initial condition of the strategies. At each section (divergent and convergent), the plot of the left side shows its evolutionary trajectory [$f_{p0}$=vertical axis, $f_{p1}$=horizontal axis, dashed lines = coefficient of variation, dots = averaged value of the trait according to time step from the initial points; oranges: $f_{p0}$=0.3, $f_{p1}$=0.7 and blues: $f_{p0}$=0.7, $f_{p1}$=0.7]. The plot of the upper right side represents the failure (vertical axis) according to time step (horizontal axis) with two different initial points of the strategies (oranges: $f_{p0}$=0.3, $f_{p1}$=0.7, blues: $f_{p0}$=0.7, $f_{p1}$=0.7). The plot of the lower right side shows the two simulations' coefficients of variation (cv = dashed lines) of the $f_p$ values separately and their centrality (C = dotted lines) which is applied from the initialized random network property. Initialized parameters of the simulation is: nodes $n$=100, connection $p$=0.9, capital $c$=1, $f_m$=0.1, $s$=100, $p_{p,max}=1$, $c_{p,1/2}$=0.5, $p_r$=0.9, $p_e$=0.05, $\mu$=0.0, $\sigma$=0.02, $p_n$=0.001, $p_l$=0.1, $t$=1000.

The case in Figure 4 seems to show that we are observing convergence or divergence to an evolutionary trajectory in the trait space ($f_{p0}$, $f_{p1}$). This makes the evolutionary phase portrait on the right-hand side meaningful and interesting: even though each new random seed produces a different evolutionary trajectory, at least the time series of outcome appears to converge. Notice we could recognize that the convergence was not every time but when it did not occur (the evolutionary phase portrait on the left-hand side), there was usually such a big variance (or fluctuation). Thus, in case of when the convergence happens, we continue to use a coefficient of variations for the horizontal and vertical lines, indicating the degrees of polymorphism in the evolutionary phase portraits as we observed the plot of the right-hand side, and more initial conditions are added to the evolutionary phase portraits.

In Figure 5.1, we have followed the same non-evolutionary part of this model [seven parameters: nodes ($n$), connection probability (p), maintenance ($f_m$), propagation probability at each node ($p_n$), propagation through each link ($p_l$), protection maximum ($p_{p,max}$), reference point ($c_{p,1/2}$)] except for the propagation probability through link ($p_l$) as control parameter. We have also used the same evolutionary part of this model [four parameter: imitation probability ($p_r$), selection intensity ($s$), exploration probability ($p_e$), and normally distributed increment of the exploration ($\sigma$)] but reduced more of the exploration rate as (as $p_e$=0.05) and its normally distributed increment (as $\sigma$=0.015) with some additional simulations from other initial conditions and time steps (see the Initialize parameters in Figure 5.1).

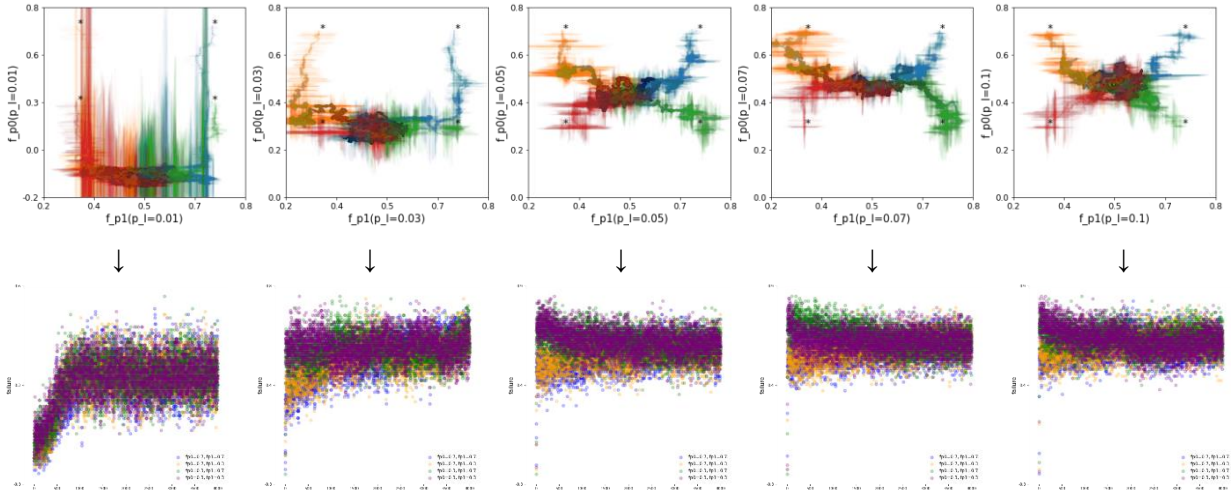

Figure 5.1: Patterns of evolution phenotypes obtained from the different starting point of the strategies with $p_l$ [Initialize parameters: nodes $n$=100, connection $p$=0.9, capital $c$=1, $f_m$=0.1, $s$=100, $p_{p,max}$=1, $c_{p,1/2}$=0.1, $p_r$=0.9, $p_e$=0.05, $\mu$=0.0, $\sigma$=0.0125, $p_n$=0.1, $p_l$=0.01 ~ 0.1, $t$=4000]. The plots show their evolutionary trajectories with five observations including its variations [$f_{p0}$=vertical axis, $f_{p1}$=horizontal axis, dashed lines = coefficient of variation, dots = averaged value of the trait according to time step from the initial points; oranges: $f_{p0}$=0.3, $f_{p1}$=0.7, blues: $f_{p0}$=0.7, $f_{p1}$=0.7, greens: $f_{p0}$=0.7, $f_{p1}$=0.3, and purples: $f_{p0}$=0.3, $f_{p1}$=0.3]. The plots of the bottom side represent the failure (vertical axis) according to time step (horizontal axis) with the different initial points of the strategies.

The coevolutionary characteristics of fixed points observed here are based on the communities with 0 < S < N (S=strategies, N=population) with a couple of different initial points. The phase portrait shows how the dynamical stability of species ($f_{p0}$ and $f_{p1}$) evolves in that space generated by the two-dimensional systems. From the observation above we draw the basic kinds of intuitions with respect to the following statements. First, for coevolutionary communities with S > 1, comprising several initial conditions, the notion of convergence, which proved useful in the classification of fixed points for the general identification of dynamical stability as demonstrated; (i) If each strategy (=species) is convergent, the fixed point might be an evolutionary attractor. (ii) If one strategy is convergent and the other divergent, the fixed point might be an evolutionary repellor. (iii) In all cased not covered by (i) and (ii), local stability of the fixed point can be turned just by varying the ratio of the evolutionary rate coefficients. Next, according to the obtained coevolutionary trajectories (upper side of the Figure 5.1), the strategy values may still evolve in time, however, outcomes like failure (or capital) reach a stationary state (bottom side of the Figure 5.1) even if the trajectories from the strategies exhibit different behavior. Therefore, the presence of a convergence indicates the possibility of a stationarity of the coevolutionary process which can depend critically on detailed dynamical features of the system (payoff, failure, and strategies). Finally, depicted evolutionary trajectories can be adaptive not only in a non-evolutionary (macro scale) controller [i.e., propagation of the failure through each link of the random network ($p_l$)] but also in an evolutionary (micro scale) controller [i.e., normally distributed increment of the exploration ($f(x|\mu, \sigma^2) = \frac{1}{\sqrt{2\pi\sigma^2}} exp^{-\frac{(x-\mu)^2}{2\sigma^2}} \big| x = individual\ capital, \mu \in R = mean(location), \sigma^2 > 0 = variance(squared\ scale)$)] when the polymorphism served as constants [i.e., the case of coexistence of failure and absence of failure from the eigenvector centrality ($\lambda x$) and imitation ($p_i$) high enough]. The result also shows that stationarity is reached by running the simulations for longer, inspired by the primary feature of the systemic risk simulation. Note that to check the validity of these formulas by simulation, we realized that we should be careful about the total simulation time; which seems that the value of variable converges to the stationary sates are not the same time, depending on the initial state of the parameters (failure propagation probability) as below.

Table 1: Stationary observation with different $p_l$.

| n | 1 | 2 | 3 | 4 | 5 | 6 | 7 | 8 | 9 | 10 | 11 | 12 | 13 | 14 | 15 | 16 | 17 | 18 | 19 |
|---|---|---|---|---|---|---|---|---|---|---|---|---|---|---|---|---|---|---|---|
| $p_l = 1$ | 1.000 | 0.632 | 0.767 | 0.718 | 0.736 | 0.729 | 0.732 | 0.731 | 0.731 | 0.731 | 0.731 | 0.731 | 0.731 | 0.731 | 0.731 | 0.731 | 0.731 | 0.731 | 0.731 |
| $p_l = 0.1$ | 0.100 | 0.125 | 0.138 | 0.149 | 0.194 | 0.228 | 0.359 | 0.412 | 0.505 | 0.573 | 0.611 | 0.637 | 0.648 | 0.662 | 0.693 | 0.701 | 0.719 | 0.728 | 0.731 |

For the different initial probability of the $p_l$ (when $p_n = 0.1$, $n = 10$), more time is needed to reach the stationarity as shown in Table 1 (we used $t$=4,000 to achieve this).

Thus, we extract the state values of $f_{p0}$ and $f_{p1}$, taking the average over a suitable time interval that only contains fluctuations around the asymptotic.

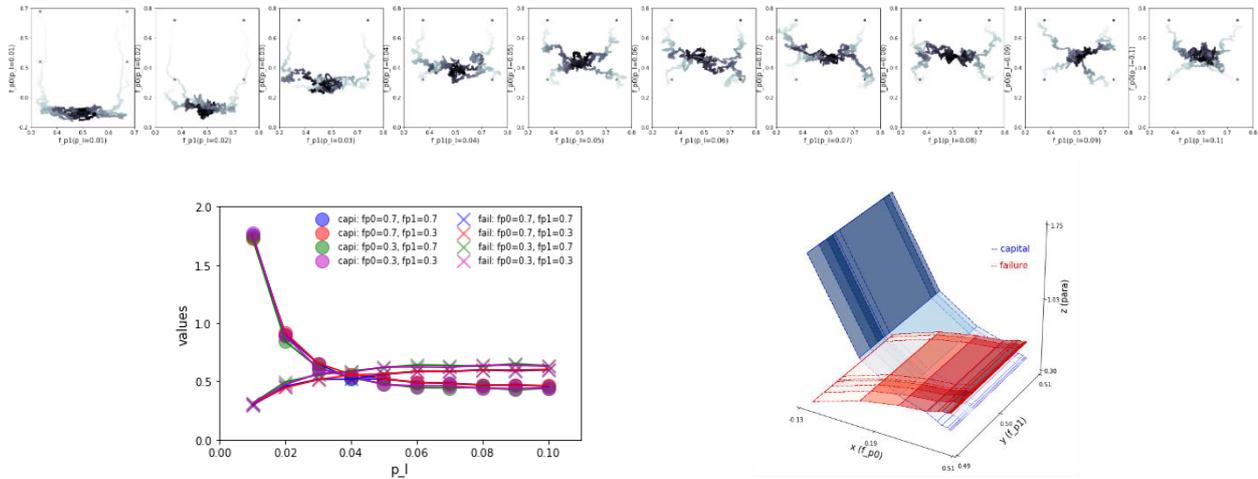

Figure 5.2: Stationary observation according to time step. The plots of the upper side show their averaged trajectories [asterisk=initial points of the strategies, strength of the color=time steps: $t$=1(weak) ~ $t$=4000(strong), dots = averaged value of the trait according to time step from the initial points; upper right: $f_{p0}$=0.7, $f_{p1}$=0.7, upper left: $f_{p0}$=0.7, $f_{p1}$=0.3, bottom right: $f_{p0}$=0.3, $f_{p1}$=0.7, and bottom left: $f_{p0}$=0.3, $f_{p1}$=0.3]. The plot of the bottom left side shows the averaged outcomes value (vertical axis) of the capital (marker=○) and the failure (marker=X) according to the $p_l$ (horizontal axis) with different initial points of the strategies (see legend). The plot of the bottom right side shows the average capital (blue) and failure (red) probability as functions of the strategies ($f_{p0}$, $f_{p1}$) with three-dimensional representation.

Table 2: Numerical result of the failure controlled by ($p_l$ obtained from the Figure 5).

| $domain(p_l)$ | 1 | 2 | 3 | 4 | 5 | 6 | 7 | 8 | 9 | 10 |
|---|---|---|---|---|---|---|---|---|---|---|

| | | | | | | | | | |
|---|---|---|---|---|---|---|---|---|---|
| *fixed mean (failure)* | 0.000 | 0.304 | 0.466 | 0.538 | 0.564 | 0.591 | 0.606 | 0.608 | 0.616 | 0.618 |
| *fixed mean (capital)* | 1.748 | 0.887 | 0.633 | 0.552 | 0.499 | 0.472 | 0.467 | 0.455 | 0.451 | 0.450 |
| *fixed mean ($f_{p0}$)* | -0.102 | 0.105 | 0.270 | 0.384 | 0.472 | 0.447 | 0.470 | 0.476 | 0.483 | 0.479 |
| *fixed mean ($f_{p1}$)* | 0.492 | 0.489 | 0.493 | 0.508 | 0.494 | 0.486 | 0.492 | 0.491 | 0.509 | 0.506 |

Although, the existence of an evolutionary attractor seems to be convergence in a certain point, we may not be able to say this is obvious because the trajectories do not have the same end point (upper side of the Figure 5.2). However, their coevolutionary trajectory fluctuations around a fixed mean value converge, at the same time, averaged capital and failure proportion reach a stationary state as the lower side of Figure 5.2 in terms of their extracted time series ($t=1\sim 4{,}000$). Concerning the outcome according to the time series, we suggest the numerical tests to explain the results of the stationarity obtained above.

***Proposition***: Stationarity implies that if we shift time by an arbitrary finite interval, then process properties do not change (they neither increase nor decay). For any initial distribution, the process will converge to the stationary probability. Simply potential dynamics as follows; $\frac{d}{dt}x = h(x) + \sqrt{2\varrho}\Gamma(t)$, $h(x) = -\frac{d}{dx}v(x)$. Where, $h(x)$ is force, and $\varrho$ is sigma. By stationarity, the result means that all variables are almost constant with a fluctuation force around their mean value. We considered that a system with a potential that exhibiting two state levels $A$ and $B$ defined by the failure simply yields; $\begin{pmatrix}p_A\\p_B\end{pmatrix} = \begin{pmatrix}p_A(n+1)\\p_B(n+1)\end{pmatrix} = \begin{pmatrix}\alpha & \beta\\1-\alpha & 1-\beta\end{pmatrix}\begin{pmatrix}p_A\\p_B\end{pmatrix}$, $p_A = \alpha * p_A + \beta * p_B$, $p_B = (1-\alpha)p_A + (1-\beta)p_A$. Alternatively, it can be written as $\begin{pmatrix}p_A(n+1)\\p_B(n+1)\end{pmatrix} = \begin{pmatrix}1-\gamma & \beta\\\gamma & 1-\beta\end{pmatrix}\begin{pmatrix}p_A(n)\\p_B(n)\end{pmatrix}$. With $\gamma = 1-\alpha$, in this case, the stationary probability is given by (i.e., $\gamma = 0.367, \beta = 1$, reflecting one of the simulation results of the failure) $p_A = \frac{\beta}{1-\alpha+\beta} = \frac{1}{1+\gamma/\beta} = \frac{1}{1+0.368/1} = \frac{1}{1.368} = 0.731$, $p_B = 1 - p_A = \frac{1}{1+\beta/\gamma} = \frac{1}{1+1/0.368} = \frac{1}{3.212} = 0.269$. Given that $p_A(n) = 0.731$, $p_B(n) = 0.269$, simulation results mean that the state ($p_A$) [and the other state ($p_B$)] probabilities remain constant (convergence on average).

***Proofs of the failure***: Let us now look at the frequency of failed agents in which we observed stationarity. The average number of failed agents denoted by ($N_f$) and the fraction of failed agents denoted by ($f$) remains constant. Suppose that at a given time, there are failed nodes ($N_f$) and not failed nodes ($N - N_f$) respectively. The number of not failed nodes at $t+1$ is given that $p_A(n) = N_f = 0.731$; $N_f + p_p(N - N_f) = 0.731 + 0.5(1 - 0.731) = 0.865 = N_{Nof}$. Where $p_p$ denotes protection probability (arbitrarily designated as 0.5 only for this numerical calculation); conversely, the number of failed nodes is $(1 - p_p)(N - N_f) = (1 - 0.5)(1 - 0.731) = 0.135 = N_f$. Then, if we impose the failure potential can propagate through each link, and the propagation probability, $p_l$, is less than 1 (when originate probability $p_n$ = constant), this equation should be modified by simply replacing the $p_p$ by $p'_p = 1 - (1 - p_p)(1 - (1 - p_l p_{ER})^{N_f}$. For example, when we arbitrary consider the protection $p_p = 0.5$ and failure propagation ($p_l$) through the random network ($p_{ER}$) to be high enough $p_l p_{ER} = 0.9$, the protection influenced by the failure propagation $p'_p$ becomes small because $1 - (1 - 0.5)(1 - (1 - 0.9)^{N_f} = 0.09$. On the other hand, if there is weak failure propagation like $p_l p_{ER} = 0.1$, the protection influenced by the failure propagation $p'_p$ becomes large because $1 - (1 - 0.5)(1 - (1 - 0.1)^{N_f} = 0.95$.

***Proofs of the capital***: Now we calculate the average value of capital at the stationary state. First, we consider the case $p_l p_{ER} = 1$. It can easily be seen that the average value of capital in stationary states following $1 + p_p(1 - f_p - f_m)c = c$ where $c$ is the average value of capital among individuals. Combining Eq. [updated capital: $1 + p_p(1 - f_p - f_m)c = c$], protection probability: $p_p = p_{p,max}/(1 + c_{p,1/2}/(f_p c))$. Then, if we consider $p_l < 1$, the protection probability $p_p$ in Eq. [$1 + p_p(1 - f_p - f_m)c = c$] should be replaced by $p'_p = 1 - (1 - p_p)(1 - (1 - p_l p_{ER})^{N_f})$. Thus, as we discovered that the protection probability is dependent on the failure propagation ($p_l$) through the random network ($p_{ER}$), capital function should be modified as follows as well; as follows: $c = 1 + (1 - (1 - p_p)(1 - (1 - p_l p_{ER})^{N_f}))(1 - f_p - f_m)c$. This shows that when the replaced protection by $p'_p$ becomes small $[1 - (1 - 0.5)(1 - (1 - 0.9))^{N_f}) = 0.09]$ because of the high enough $p_l p_{ER} = 0.9$, the capital at stationary is going to be less than in the case of $p'_p$ which becomes large because of the weak failure propagation $p_l p_{ER} = 0.1$ [i.e., capital $c = 1 + p'_p(1 - f_p - f_m)c$ =1.036 = 1+0.09(1-0.5-0.1)c < capital $c = 1 + p'_p(1 - f_p - f_m)c$ = 1.38 = 1+0.95(1-0.5-0.1)c].

The intuition from the results at the stationarity is that even if (i) the strategy of $f_{p0}$ and the strategy of $f_{p1}$ behaviors are not the same: as we can notice in the trajectories representation in Figure 5.2, while the strategy of $f_{p0}$ increase their states according to the controlled parameter value of $p_l$ in the applied random network $p_{ER}$, the other strategy of $f_{p1}$ multiplied by the eigenvector centrality ($C$) is not sensitive about their influence, (ii) the failure and capital show clear correlation according to the controlled parameter value change as we observed in Figure 2 and 3. Moreover, (iii) its impact seems to be much more significant in the early stage of the state following called a power law.

***Generalization of the obtained outcomes***; This characteristic is of course not just restricted to this simulation controlled by the $p_l$ but is given the more general name of the network effect; every time someone links to a particular node on a network, it makes it that bit more likely that someone else will also; $p_{ER}$ determined by their connection probability. We expend our interest specifically in the $p_c$ [when propagation probability ($p_l = 0.1$) constant] because the interpreted function of $p'_p = 1 - (1 - p_p)(1 - (1 - p_l p_{ER})^{N_f})$ was not only dependent on the failure propagation ($p_l$) but also through the random network ($p_{ER}$) property [from the eigenvector centrality ($\lambda x$) controlled by the connection probability ($p_c$)]. And we could observe a quite similar trend of their outcomes as we observed in the simulation results controlled by the $p_l$.

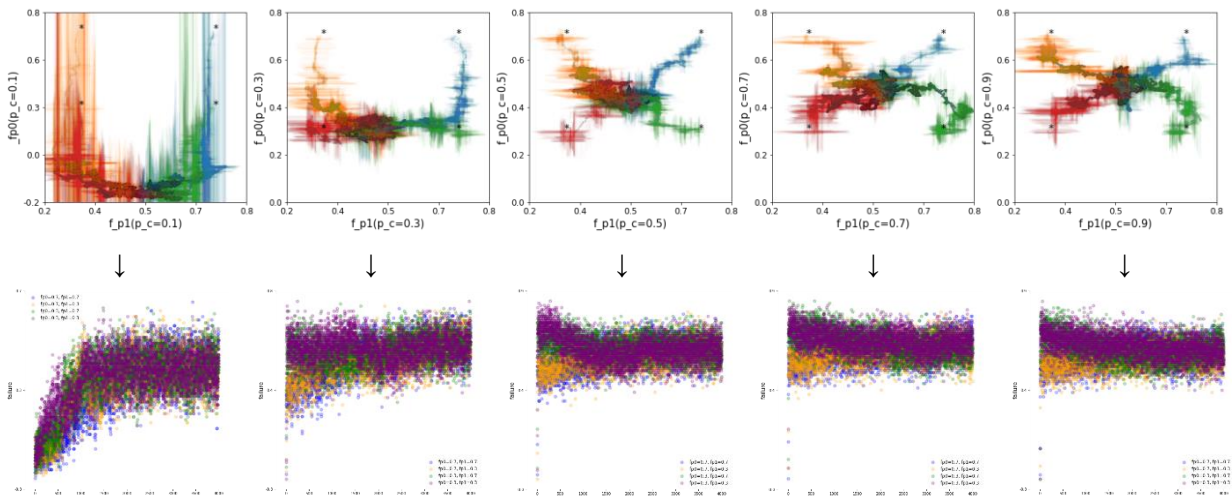

Figure 6.1: Patterns of evolution phenotypes obtained from the different starting point of the strategies with $p_c$ [Initialize parameters: nodes $n$=100, connection $p$=0.1~0.9, capital $c$=1, $f_m$=0.1, $s$=100, $p_{p,max}$=1, $c_{p,1/2}$=0.1, $p_r$=0.9, $p_e$=0.05, $\mu$=0.0, $\sigma$=0.0125, $p_n$=0.1, $p_l$=0.1, $t$=4000]. The plots show their evolutionary trajectories with five observations including its variations [$f_{p0}$=vertical axis, $f_{p1}$=horizontal axis, dashed lines = coefficient of variation, dots = averaged value of the trait according to time step from the initial points; oranges: $f_{p0}$=0.3, $f_{p1}$=0.7, blues: $f_{p0}$=0.7, $f_{p1}$=0.7, greens: $f_{p0}$=0.7, $f_{p1}$=0.3, and purples: $f_{p0}$=0.3, $f_{p1}$=0.3]. The plots of the bottom side represent the failure (vertical axis) according to time step (horizontal axis) with the different initial points of the strategies.

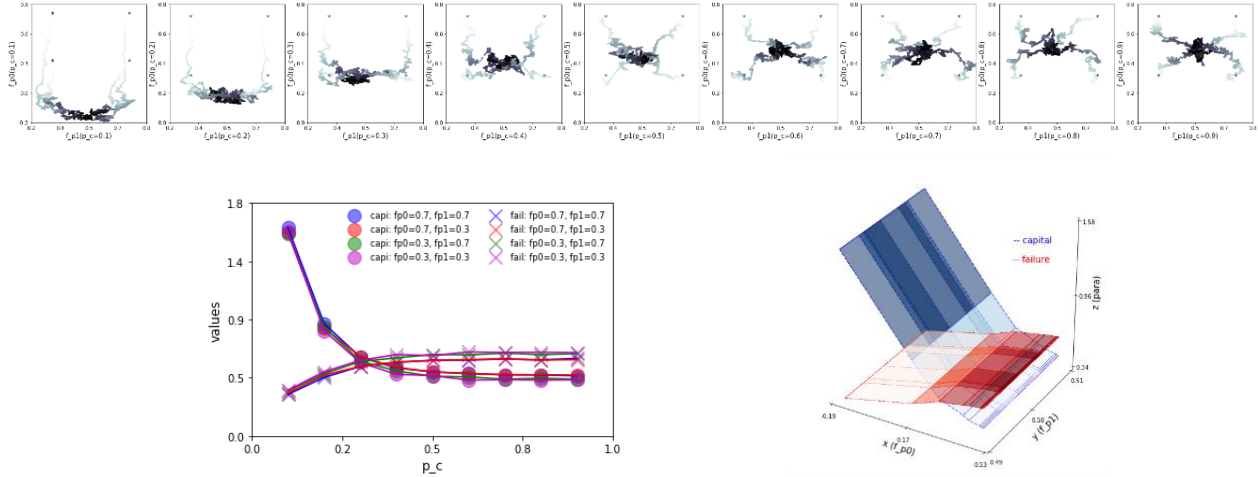

Figure 6.2: Stationary observation according to time step. The plots of the upper side show their averaged trajectories [asterisk=initial points of the strategies, strength of the color=time steps: $t$=1(weak) ~ $t$=4000(strong), dots = averaged value of the trait according to time step from the initial points; upper right: $f_{p0}$=0.7, $f_{p1}$=0.7, upper left: $f_{p0}$=0.7, $f_{p1}$=0.3, bottom right: $f_{p0}$=0.3, $f_{p1}$=0.7, and bottom left: $f_{p0}$=0.3, $f_{p1}$=0.3]. The plot of the bottom left side shows the averaged outcomes value (vertical axis) of the capital (marker=○) and the failure (marker=X) according to the $p_c$ (horizontal axis) with different initial points of the strategies (see legend). The plot of the bottom right side shows the average capital (blue) and failure (red) probability as functions of the strategies ($f_{p0}$, $f_{p1}$) with three-dimensional representation.

Table 3: Numerical result of the failure controlled by ($p_c$ obtained from the Figure 6).

| $domain(p_c)$ | 1 | 2 | 3 | 4 | 5 | 6 | 7 | 8 | 9 |
|---|---|---|---|---|---|---|---|---|---|
| $fixed\ mean\ (failure)$ | 0.000 | 0.338 | 0.474 | 0.558 | 0.592 | 0.604 | 0.612 | 0.612 | 0.617 |
| $fixed\ mean\ (capital)$ | 1.582 | 0.839 | 0.591 | 0.508 | 0.479 | 0.462 | 0.455 | 0.454 | 0.451 |
| $fixed\ mean\ (f_{p0})$ | -0.155 | 0.170 | 0.291 | 0.405 | 0.432 | 0.481 | 0.487 | 0.488 | 0.492 |
| $fixed\ mean\ (f_{p1})$ | 0.497 | 0.500 | 0.492 | 0.497 | 0.510 | 0.504 | 0.506 | 0.496 | 0.506 |

Figure 6 shows that the scatter plot which is constructed by the connection probability ($p_c$) do obey strong power-law relationships like the previous observation $p_l$ (see the fixed mean value of the failure in Table 2 and Table 3). In order to interpret the potential state variable (placed as a probability) according to the control parameters of $p_c$ compared to the $p_l$, let us use a simple way [going back to the primary feature of the failure which mentioned in the section of *Cascading failure in the network* ($R = K/S$)] to specify this detail with numerical test. Intuitively, we assume that the parameter value of $p_l$ or $p_c$ can be considered as the main variable ($x = p_l p_{ER}$) in this case and cannot equal zero since the [$p_l \in (0,1)$, $p_c \in (0,1)$]. Yielded function would be [$f(x) = k/x,\ x \neq 0$] as simply as possible to explain what happens. We assumed the denominator of the fraction as domain [greater than zero ($x > 0$)] which is from the probability of the failure propagate through each link ($p_l$ in case of the Figure 5), at the same time from the probability of the connection in the random network ($p_{ER}$ determined by $p_c$ in case of the Figure 6). If we keep the nominator of the fraction as constant ($k$) which can be from an inherited fitness for every individual at the time, and plus one more value in that each agent receives one unit of payoff in each time steps ($c = 1$: corresponding to the model's *payoff dynamics*), such that function will be $[f'(x) = -(k/x) + c]$.

Table 4: Numerical result of the function $[f'(x)]$.

| $domain(x)$ | 1 | 2 | 3 | 4 | 5 | 6 | 7 | 8 | 9 | 10 |
|---|---|---|---|---|---|---|---|---|---|---|
| $f'(x)$ | 0.000 | 0.500 | 0.666 | 0.750 | 0.800 | 0.833 | 0.857 | 0.875 | 0.888 | 0.900 |

In what follows, such a range $[f'(x)]$ will be decided by the failure propagate through each link in the random network ($p_l p_{ER}$) which was determined by both parameters ($p_l$ and $p_c$) with the same weight. We simply proved this numerical trend about the defined protection probability by $p'_p = 1 - (1 - p_p)(1 - (1 - p_l p_{ER})^{N_f})$ must be the similar by the $p_l$ [when the $p_{ER}$ = constant (Table 2)] as well as by the $p_c$ [when the $p_l$ = constant (Table 3)]. Thus, failure influences in these simulations are almost identical as we obtained in the Table 2.1 according to these parameters change including where increase or decrease occurs following a nonlinear curvature of the power law.

This interpretation helps to illustrate the dynamics behind how these propagations through the system can move or develop in a particular direction as many real-world networks (such as finance, supply chain, and disease etc.) have proven this power law relationship between size and quantity. We should note here again applied potential in here can go in a short period of time which we might cite as a bias or rationality. The inspiration from this observation and interpretation is to get a sense of the qualitatively different nature of propagation within systems that the failure is not just growth or decay, but due to the network effect over time, there is also another rate that is itself increasing the risk. This statistical core of the phenomena played an additioanl part in the underlying phase transitions across a wide range of the system in this model.

**Part4: Summary of the results**

We tested the model we created by conducting simulations on random network graphs generated as follows: the network property for each occupies a vertex (drawn as a node); the edge (illustrated as a line connecting two vertices) marks the nearby sites where a reproducing individual can place an offspring. With respect to the property of this network, the rest of the graph is organized as follows: the size of the node denotes the number of edges incident on a node (drawn as a degree); the width of the edges represents an influence based on that connection (drawn as an eigenvector centrality) which randomly chooses each neighbor. The potential of the systemic risk then propagates through failure dynamics as does the artificially designated probability with the initial event of the invading failure. Proper protection against the systemic risk with the system components evolving heuristically through strategy dynamics (social learning and exploration) as a potential for absence of failure.

Notice how it all came about and what the results mean to us. The model mechanism assumes that any individual that has a failure would devastate the whole system. Every one of them is in the situation of being what is called systemically essential, and all these will systemically cascade. Let us specify what the simulation mechanism would say to reflect this point. First, there is a (a) contagion. If one individual (=node) fails because of its interconnected relationship in a network (for example, an industry) its failure will have an impact on other individuals in the industry. There is thus a contagion type of effect (reflected by links). Second, there is a (b) concentration. In the industry, even it only one player or small entity with a massive potential is more extensive than all the others, there will be a significant concentration (reflected by the eigenvector centrality). Third, there is a (c) context, namely, what usually happens to an individual's function in an ordinarily operable environment; if the circumstance takes a turn for the worse, perhaps so too will all other individuals, as all of these individuals and institutions are in context (reflected by social learning and imitation). Any individual in the industry functions in the same way. It also takes the same kinds of risks as others take. Thus, if one goes down, they all go down. We observe a significant correlation between capital and failure at both the micro- and macro-scale, which is another critical fact. Note that if an individual without protection goes in the direction eliminating the risk through investment, every single individual in that context will go to fail.

Another potentially important factor in these results is a cultural evolution and the dilemma it poses for imitation and exploration through social learning. The random probability controls the influence, and the impact depends on the choice between arbitrary designated strategies ($f_{p0}$ and $f_{p1}$) among the individuals through social learning and exploration. The output suggests that failure might be a critical driver for bias with a given centrality, social learning, and exploration. Evolutionary heuristics remains the dominant measurable unit of credit in any dynamics. Given the reliance of most plausible principles on the network, the dynamics of accumulated strategies has been scrutinized by generations of explorations. From fundamental work connected with this model, we know that the normally distributed increment of explorations with respect to protection level is highly skewed. Many studies are never proved, and this uneven exploration distribution is a robust, emergent property of the dynamics of the systemic risk propagation. This means that we can compare the impact of protection level biased in different strategies by looking at their relative exploration rates.

In Figure 5, we have used the same parameters to picture the combined dynamics of trait substitution sequences in two coevolving strategies originating from different initial conditions. At any moment in time, two coefficients of variation were calculated from the simulations by dividing the standard deviations of $f_{p0}$ and $f_{p1}$ by their respective means. We plot these as functions of time, which yield two curves for one initial condition in the plot, one for $f_{p0}$ and the other for $f_{p1}$, with time on the horizontal axis and the coefficients of variation on the vertical axis. We added some additional simulations from other initial conditions. The main point of interest here is whether they all do converge to the stationarity. Of course, each new random seed produces a different evolutionary trajectory, as we mentioned above, but as our previous time series of failure proportions appear to converge, so the update with more initial conditions has converged to an asymptotic value in a certain parameter setting. The results show that individuals outcomes (failure and bias) at the stationarity follows a power law whose tails can be called a preferential attachment in network dynamics [36].

# Discussion

We present a simple general model to quantify the protection that will be used to mitigate systemic risk, with results as follows: First, (i) the model introduces the nature of network property. Then (ii) it suggests a prototypical failure impact which is needed for projecting the common risk exposure onto the set of individuals. In a last step, (iii) the model implies how protection can be applied to the network; then (iv) it simultaneously combines theses coevolutionary features. As detailed above in Methods, the present model treated a couple of results as factors in a fully undirected random design specified in the model structure and the dynamics of these simulations together with corresponding diversifications. Based on the simple set of property, the observations from this model highlight the fact that the probability describes the portion of protection which, from between nodes in the networks, can be characterized by how systemic risk should be coped with, rather than it being predicted by the probability of failure [37]. The broad spectrum of emergent behavior encapsulated in networks that have connectedness and spreading may be explained as follows.

**Systemic risk**: First, the simulations show that there is a vast variety of phenomena with random connections that arbitrarily act as robust amplifiers for the failure initialization. Many of those original properties are structurally simple (specifically, there are certain subset of vertices that we could call a small world because of their topology) but also strong (due to the influence based on those connections), and these could be realizable in other network structures such as regular and cycle in a fixation time of mutants [38]. The result provides an explicit procedure for their properties, proving the existence of those structures. The arrangement guarantees that, with high probability, the fundamental properties or influences spread along a branch corresponding to their universal characteristics at the macro-scale. The connection of all edges is then intensified so that the spreading repeatedly invades, and eventually failure spreads through the links one-by-one to all the branches until a cascade occurs at the micro-scale. Intuitively, the degree assignment creates a sense of primary flow, directed toward the initial failure which demonstrates that once the failure reaches the high centrality, the agents are highly likely to persist in invading more neighbors. Thus, if we know the starting links and finishing links of the node, we might be able to assess the risk in the network system. We can suggest that the risk emerges through common conditions caused by the relations of the node. The intuition is that, because we did estimate with the immutable characteristics, if we remove links on the node to the system, the specification must be added to the system, and the value of links will be equivalent to the systemic risk, as we saw in the simulation result. We observed that the underlying networks had created random pathways along which failure events can spread rapidly and globally, which we might call systemic risks. In this perspective, we argue that systemic failures are consequences of the highly interconnected systems (connection probability) and networked risks (failure probability) that individuals have created. Such interdependencies will inevitably get out of control, and a network-oriented view can understand the instabilities.

**Protection potential**: Second, the simulation model explicitly regulates the potential of the protection to the systemic risk by interconnected dynamics. As can be seen in the result part of 1 (a *feature of systemic risk*), it is not possible to predict or control the potential for catastrophic failure even though all the information may be embedded in the system at the macro-scale (=universal application of the parameters). Such problems might be solved by suitable management applications [and proper (re)design of the structure]. We constructed a suite of plausible dynamics, decentralized bottom-up mechanisms, by establishing appropriate 'rules of the interaction', within which the system components can self-organize, including mechanisms ensuring rule compliance (vectorized micro-scale implementation). Evolutionary dynamics, for example, can often promote a well-balanced situation with respect to the interactions. When the investment into protection is weak (low investment), a pattern of strong systemic risk emerges as agent failure in networked conditions.

In contrast, when the investment into protection is strong (high investment), a pattern of protection emerges, with a little diversification against all challenges. The results cast light on the modes of propagation. Observed contagion and persistence patterns should be viewed not as a direct causal link, but rather as due to an accumulated rational driven by interconnectedness [20]. Moreover, the failure potential is also due to time delays after an individual is officially failed [39]. Governments or politics are often reluctant to resolve insolvent institutes (i.e., banks, firms, supply chains etc.) and they permit the individuals to continue operating amid negative effects. The length of these delay causes not only increased failure but also capital losses, which is more likely for insolvent or near-insolvent individuals, leading to reductions in network welfare. These delays may at times stretch to many others in the network, increasing the fragility and the probability of failure [40]. Many pieces of evidence in most governing systems do indeed suggest that if troubled individuals could be estimated before the value of their capital turned negative, institutions, based on their risk potential would be permitted to weed out the inefficient or unfortunate individuals to protect against more serious adverse effects [41]. This reinforces the importance of resolving individuals as quickly as possible and developing faster procedures for certifying protected individuals by providing immediate interventions. On the other hand, prompt corrective actions may increase the willingness to supply such protection to reduce the chances of systemic risk [42]. Given the evolutionary mechanisms implemented in this simulation model, we observed the evolutionary response many times in order to obtain a critical value for the plausible protection potential. We demonstrated that although the structures have high failure potential in terms of the systemic risk, the function of the interconnection turns them into weak amplifiers, where profitable investment with high protection.

**Strategy dynamics**: Third, the simulation shows that cultural evolutions are key features of investment for protection against the spread of failure. Namely, the results confirmed that without either exploration through social learning, no strategy is a robust amplifier under the propagation of failure and no investment is a high enough one to achieve an absence of failure under the contagion. We notice that clear potential, such as cultural evolution, can be turned into arbitrarily powerful amplifiers, and the results of experiments on the strategy dynamic vary the fitness advantages for the protection. Depending on the investment into protection through the bottom-up dynamics, it is possible to find individuals' strong bias [43], who alter their competitive strategy [14], that have a large potential for protection, and that reconcile the networked agent's broad range of the systemic risk values as basics of their interconnected interactions. In particular, inspired by the plausible scenarios which we presented, we focusd on the parameter of ($f_p c$) with more individuals and time steps. The results give us some insights into the fact that the regime in the strongly centralized but weakly with interactions between individuals might be not recommendable in the current model mechanism, and that biases might be mitigated by exploration if there is enough social learning. Shifting away from the existing dominant potential, norms, and routines can be productive by leading to novel trends with high impact. Evidence shows that biases between parameters occur at very early stages— hence, the exploration (social learning) probability and rates that could be another crucial factor offering resources to violate expectations and leads to novel trends with high impact. Following the observations, we can assume that successful outcomes require there to be a delicate balance across the essential components (structure, centrality, social learning, and exploration). To be valuable, individuals and organizations might seek legitimacy to reduce their perceived risk potential by associating with this finding. Although many questions remain regarding how different network structure, governance, and content evolve and interact over time, resources help to gauge the underlying potential. Observation in this simulation may lead to a subsequent beneficial network structure being engendered in social relations that is jointly supported by governance.

**Coevolutionary features**: Finally, there has been some interest in the general question as to whether the phenotypes evolve as an evolutionarily stable strategy [44, 45]. According to researcher [46], the interaction between strategies prevents the attainment of a converging point, such that there is a continuous evolutionary change in their phenotypes [47]. Inspired by the investigation process of this model, we thus utilized the three dynamics of coevolution (payoff, failure, and strategy) to investigate the variety of possible evolutionary traits in a random network. In particular, we focused on the potential for stationarity and observed that this mode of coevolution is a feasible outcome. In Figure 5, it is seen that the adaptive trait values tend to a converging trend, as, once this is reached, no further fluctuations occur around the fixed mean value. This finding corroborates speculations put forward regarding the necessity for such interactions to motivate a variety of phenotypic coevolution. Such dynamics are interpreted as indicating the continuous deterioration of a strategy's environment owing to the continual evolution of other strategies [48]. Analysis of these interpretations suggests that, through evolution, the phenotypes could either tend to convergent or to divergent asymptotic states. We have seen in the results that in a random network a variety of evolutionary outcomes is possible. According to the framework established here, we ensure that the process of directing evolution is driven explicitly by the interactions of different phenotype, these being the events that arise from encounters with other individuals [49], as opposed to the constant birth and death events. We assume that variation is created by an interconnected evolutionary process; to keep the analysis tractable, we envisage that the various functions could be used for this purpose; the main function describing the effect of that individual is likely to show some degree of specialization in the features of stationarity. In the case of the simulations, we can immediately infer from observations that there is a region in the monomorphic trait space where the strategies can coexist [50]. Such volatility can be governed not only through quantified non-evolutionary part but also by identifying the evolutionary part of the strategies, such as individuals' imitation and exploration as they aim to increase their protection [51]. Furthermore, emerging phenomena measured by the interconnected contributions obey that power law like other network systems follow [52]—which a mathematical formula fits well to plausibly explain the likely outcomes [36], and the probability distributions based on the individual characteristics can play a prominent role in discourses about their potentials [53].

**Concluding remarks:** Research on systemic risk has yielded many remarkable findings. Our observations indicated that successful outcomes do indeed seem to require a balance across objects. From empirical results, there is also a consensus that embeddedness in a network of interrelations matters for the network's payoff and performance [54]. In contrast, only a partially oriented confirmation was conducted for a network dynamic, although the report alone does not tell the full story [55]. Critically scanned uncertainty needs to be addressed to answer questions as to how content network relationships, governance, and structure emerge over time [56]. With a random network-agent model, we notice that the proposed "protection dynamics" leads to a restructuring of the systemic risk that is practically free of failure. The principle, being simple but fundamental, does not ordinarily change across individual; hence, we suggest that this is a straightforward and easy model which can be used for finding invariant properties and provide an example so that it can be used when needed. Rules and process implemented in this study would provide a fresh way for decision-makers (or social planner) to gain a better perspective in the dynamics of systemic risk. With a simple network behind this model consisting of two types of agents, an interesting direction for future research will be motivated by whether comparable results can be achieved for systemic-risk. We could enhance the prospects of systemic risk as a whole to more effectively address its related problems [12] by attempting to eradicate an infection on the optimal actions of a decision making.

# Supplement information of the Model

## Mathematical description of the mechanisms

In computation, there are rules of thumb that we can implement into an algorithm to help it solve many problems. These usually do not work in every case, and we do not need them to. We need them to work for a problem to which we have devoted more effort to optimize them. One case to which we have given great attention is linear programming. The fundamental idea is that we have a matrix $A$, a vector $B$, and we want to find vectors such that i.e., $Ax$ is less than or equal to $B$;

$$\{x \in \mathbb{R}^n | Ax \leq B\}, \qquad \mathbb{R}^n = n\ dimensioanl\ set\ of\ real\ numbers$$

Here, each entry of vector $A$ is the corresponding entry of vector $B$, which shows up all the time in optimization. The heuristic here is if we have a problem that we really want to solve because of the amount of effort that people have put into it, we could try reducing it to one of these problems and plugging it into the solvers that exist. Instead of giving ourselves a hard task, we reduce our problem to find a reduction in programing in which the existing algorithms can work well, such that linear programing can take advantage about many case complexities of algorithms.

***Systemic risk in the random network:*** We put forward the proposition regarding the emerging protection and its potential far-reaching impacts on agented-network under the basic operating principle mentioned above. We therefore suggest the following steps for modelling. First, (*network properties*) when constructing the basic data structure, the functionalities of the mechanism begins with a specific undirected relationship between agents. Next, (*primary risk influence*) using a parameter to evaluate the impact of risk for the networked agents, the influence of primary risk is estimated along the structure as a general failure property. Finally, (*protection against systemic risk*) embedding a protection dynamic by emphasizing the role of payoff, failure, and strategy dynamics.

The structure created (network) is undirected and starts by adding edges between pairs of nodes one at a time randomly (Erdös-Renyi graph). For example, 4 nodes (vertices) added possible edges (line);

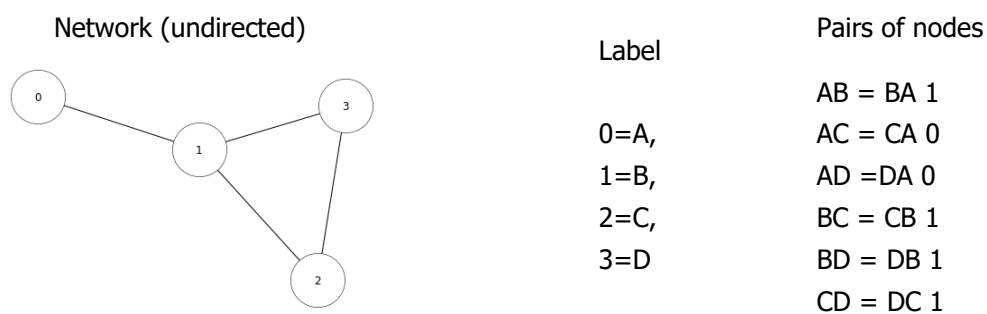

| Network (undirected) | Label | Pairs of nodes |
|---|---|---|
| | | AB = BA 1 |
| | 0=A, | AC = CA 0 |
| | 1=B, | AD =DA 0 |
| | 2=C, | BC = CB 1 |
| | 3=D | BD = DB 1 |
| | | CD = DC 1 |

Describing a graph by an adjacency matrix: If we start with the idea of an adjacency matrix, we think of the rows of the matrix as well as the columns of the matrix to be labelled by the vertices (nodes), so here we have 1, 2, 3, & 4 vertices. The actual labelling that we give can be anything but let us call our adjacency

"matrix A" and this is the actual definition of our matrix. An entry in row $m$ column $n$ will be equal to either 1 or 0. It will be equal to 1 if there is an edge between $m$ and $n$ and if $m$ is not connected to $n$, it will be zero

$$Am \times n = \begin{matrix} & \begin{matrix} 1 & 2 & 3 & 4 \end{matrix} \\ \begin{matrix} 1 \\ 2 \\ 3 \\ 4 \end{matrix} & \begin{bmatrix} & & & \\ & & & \\ & & & \\ & & & \end{bmatrix} \end{matrix}, \quad A = \begin{cases} 1 & if\ mn\ \in edge \\ 0 & otherwise \end{cases}$$

The diagonal elements of the adjacency matrix are zero for graphs without loops.

$$A = \begin{matrix} & \begin{matrix} 1 & 2 & 3 & 4 \end{matrix} \\ \begin{matrix} 1 \\ 2 \\ 3 \\ 4 \end{matrix} & \begin{bmatrix} 0 & 1 & 0 & 0 \\ & & & \\ & & & \\ & & & \end{bmatrix} \end{matrix}, \quad A = \begin{matrix} & \begin{matrix} 1 & 2 & 3 & 4 \end{matrix} \\ \begin{matrix} 1 \\ 2 \\ 3 \\ 4 \end{matrix} & \begin{bmatrix} 0 & 1 & 0 & 0 \\ & 0 & & \\ & & 0 & \\ & & & 0 \end{bmatrix} \end{matrix}, \quad A = \begin{matrix} & \begin{matrix} 1 & 2 & 3 & 4 \end{matrix} \\ \begin{matrix} 1 \\ 2 \\ 3 \\ 4 \end{matrix} & \begin{bmatrix} 0 & 1 & 0 & 0 \\ 1 & 0 & 1 & 1 \\ 0 & 1 & 0 & 1 \\ 0 & 1 & 1 & 0 \end{bmatrix} \end{matrix}$$

The graph which will be discussed here will not have loops (to itself) on a single node. In an adjacency matrix, rows and columns represent vertices (nodes), so the sum on each row (column) is the degree of the related node. For the above example, the vertex 2 has 3 neighbors. For the columns of the matrix:

$$A = \begin{matrix} & \begin{matrix} 1 & 2 & 3 & 4 \end{matrix} \\ \begin{matrix} 1 \\ 2 \\ 3 \\ 4 \end{matrix} & \begin{bmatrix} 0 & 1 & 0 & 0 \\ 1 & 0 & 1 & 1 \\ 0 & 1 & 0 & 1 \\ 0 & 1 & 1 & 0 \end{bmatrix} \end{matrix}, \quad \deg(v_{2m}) = 3, \quad \deg(v_{2n}) = 3$$

If we look at the column that represents the vertex 2 we will also get 3 which represents its degree. So, keep in mind that the adjacency matrix of a graph has all the same information that is contained in the graph. Notice that the graph has been shown to us in any random way that the computer generates the visuals. If we were to run again, we would get a different picture but no matter how we run it, the same relationship between those vertices (nodes) and edges (line) creating its degrees will be the same as below;

$$\sum_{v \in V(G)} \deg(v) = \deg(v_1) + \cdots + \deg(v_n) = 2|E(G)|$$

states that twice the number of edges equal to its sum of the degrees (i.e., E=4=deg=8) as we can see in the pair of nodes. In every graph twice, the number of edges $[2|E(G)|]$ is the sum over the degree of nodes $[\deg(v_1) + \cdots + \deg(v_n)]$ or the sum over matrix elements.

$$A = \begin{bmatrix} a_{11} & a_{12} & \cdots & a_{1n} \\ a_{21} & a_{22} & \cdots & a_{2n} \\ \cdots & \cdots & \cdots & \cdots \\ a_{m1} & a_{m2} & \cdots & a_{mn} \end{bmatrix}$$

The resulting $m \times n$ matrix is obtained by $[A = G(n,p)]$, any node can be randomly exposed to the other nodes, which creates random connections. Given the collection of nodes influenced by the connection probability $[p \in (0,1)]$, the model asks what the distribution of the connection in the network is (probability of degree).

Next, to observe the process of propagation, the model uses an array (vector) as a probability of failure $[p \in (0,1)]$ with a given initially influenced nodes (1 ≤ j ≤ N) noted merely by ($p\_j$). Each node can be in one of two states; not failed or failed. All nodes are initially without failure.

Because, the previous mechanism was;

$$A = \begin{bmatrix} a_{11} & a_{12} & \cdots & a_{1n} \\ a_{21} & a_{22} & \cdots & a_{2n} \\ \cdots & \cdots & \cdots & \cdots \\ a_{m1} & a_{m2} & \cdots & a_{mn} \end{bmatrix}$$

Extending the above output given the failure dynamics becomes;

$$A = \begin{bmatrix} a_{11} & a_{12} & \cdots & a_{1n} \\ a_{21} & a_{22} & \cdots & a_{2n} \\ \cdots & \cdots & \cdots & \cdots \\ a_{m1} & a_{m2} & \cdots & a_{mn} \end{bmatrix} \begin{bmatrix} \vec{a}_1 \\ \vec{a}_2 \\ \cdots \\ \vec{a}_n \end{bmatrix} = \begin{bmatrix} a_{11}\vec{a}_1 & + & a_{12}\vec{a}_2 & + & \cdots & a_{1n}\vec{a}_n \\ a_{21}\vec{a}_1 & + & a_{22}\vec{a}_2 & + & \cdots & a_{2n}\vec{a}_n \\ \cdots & & \cdots & & \cdots & \cdots \\ a_{m1}\vec{a}_1 & + & a_{m2}\vec{a}_2 & + & \cdots & a_{mn}\vec{a}_n \end{bmatrix} = B\vec{a} = \begin{bmatrix} B\vec{a}_1 \\ B\vec{a}_2 \\ \cdots \\ B\vec{a}_n \end{bmatrix}$$

i.e., `States Matrix`

```
[[ 1.,  1.,  1.,  1.,  0.,  0.,  1.,  1.,  1.,  1.],
 [ 0.,  0.,  0.,  1.,  1.,  1.,  1.,  1.,  1.,  1.],
 [ 0.,  0.,  0.,  0.,  0.,  0.,  0.,  0.,  0.,  0.],
 [ 0.,  0.,  0.,  0.,  1.,  1.,  1.,  1.,  0.,  0.],
 [ 0.,  0.,  0.,  0.,  1.,  1.,  1.,  1.,  1.,  1.],
 [ 0.,  0.,  0.,  1.,  1.,  1.,  1.,  1.,  1.,  0.],
 [ 0.,  0.,  0.,  0.,  0.,  0.,  0.,  0.,  0.,  0.],
 [ 0.,  0.,  1.,  1.,  1.,  1.,  1.,  1.,  1.,  1.],
 [ 0.,  0.,  0.,  0.,  0.,  0.,  0.,  0.,  0.,  1.],
 [ 0.,  0.,  0.,  0.,  0.,  0.,  1.,  1.,  1.,  1.]]
```

Notice that the matrix is denoted by an $B\vec{a}$ instead of an $A$ be representation because it no longer follows the adjacency matrix representation. The $B\vec{a}$ is still going to be labelling our rows and columns via 1 and 0, the key difference is being the possibility of showing the state of each node ($m$ : 1= failure, 0 = absence of failure) according to time steps ($n$).

**Impose protection against the systemic risk:** Along with the basic intuition mentioned above, protection dynamics was applied. First of all, we break the program into sub-dynamics (payoff, failure, and strategy). The result of each sub-dynamics is saved. These sub-dynamics are trivial problems that add complexity to the dynamics. To implement each part, we use simple equations. These equations combine some previously computed variables, and newly added or computed variables. In the following example, we used the values already stored in the table to compute new variables. This technique is often called memorization.
For example;

a → store in the table
b → store in the table

$$a + b = c \rightarrow \text{lookup } a, b \rightarrow \text{compute } c$$
$$d \rightarrow \text{stored in the table}$$
$$a + d = e \rightarrow \text{lookup } a, d \rightarrow \text{compute } e$$

*Payoff dynamics*: One agent is associated with each node and is characterized by its capital and strategy as below; In each time step, each agent receives one unit of payoff, which is added to its capital $c$, of which fractions $f_m$ and $f_p$ are spent on maintenance and protection, respectively, resulting in the updated capital $1 + (1 - f_m - f_p)c$. We used an element-wise computation by using arrays for the vectorization ($\vec{v}$) instead of using a loop.

$$f_p = \vec{v}_i = \begin{bmatrix} \vec{v}_{i1} \\ \vec{v}_{i2} \\ \cdots \\ \vec{v}_{in} \end{bmatrix}, f_m = \vec{v}_{ii} \begin{bmatrix} \vec{v}_{ii1} \\ \vec{v}_{ii2} \\ \cdots \\ \vec{v}_{iin} \end{bmatrix}, \vec{v}_{ii} + (-)\vec{v}_i = \begin{bmatrix} \vec{v}_{ii1} + (-\vec{v}_{i1}) \\ \vec{v}_{ii2} + (-\vec{v}_{i2}) \\ \cdots \\ \vec{v}_{iin} + (-\vec{v}_{in}) \end{bmatrix} = \begin{bmatrix} \vec{v}_{iii1} \\ \vec{v}_{iii2} \\ \cdots \\ \vec{v}_{iiin} \end{bmatrix},$$

$$c \begin{bmatrix} \vec{v}_{iii1} \\ \vec{v}_{iii2} \\ \cdots \\ \vec{v}_{iiin} \end{bmatrix} = \begin{bmatrix} c\vec{v}_{iii1} \\ c\vec{v}_{iii2} \\ \cdots \\ c\vec{v}_{iiin} \end{bmatrix} = \begin{bmatrix} \vec{v}_1 \\ \vec{v}_2 \\ \cdots \\ \vec{v}_n \end{bmatrix}$$

Because the previous initial random network property was;

$$A \begin{bmatrix} a_{11} & a_{12} & \cdots & a_{1n} \\ a_{21} & a_{22} & \cdots & a_{2n} \\ \cdots & \cdots & \cdots & \cdots \\ a_{m1} & a_{m2} & \cdots & a_{mn} \end{bmatrix} + kA \begin{bmatrix} ka_{11} & ka_{12} & \cdots & ka_{1n} \\ ka_{21} & ka_{22} & \cdots & ka_{2n} \\ \cdots & \cdots & \cdots & \cdots \\ ka_{m1} & ka_{m2} & \cdots & ka_{mn} \end{bmatrix} = B, \quad k \in [0,1]$$

The applied output given payoff dynamics becomes;

$$B = \begin{bmatrix} b_{11} & b_{12} & \cdots & b_{1n} \\ b_{21} & b_{22} & \cdots & b_{2n} \\ \cdots & \cdots & \cdots & \cdots \\ b_{m1} & b_{m2} & \cdots & b_{mn} \end{bmatrix} \begin{bmatrix} \vec{v}_1 \\ \vec{v}_2 \\ \cdots \\ \vec{v}_n \end{bmatrix} = \begin{bmatrix} b_{11}\vec{v}_1 + b_{12}\vec{v}_2 + \cdots b_{1n}\vec{v}_n \\ b_{21}\vec{v}_1 + b_{22}\vec{v}_2 + \cdots b_{2n}\vec{v}_n \\ \cdots \cdots \cdots \cdots \\ b_{m1}\vec{v}_1 + b_{m2}\vec{v}_2 + \cdots b_{mn}\vec{v}_n \end{bmatrix} = B\vec{v} = \begin{bmatrix} B\vec{v}_1 \\ B\vec{v}_2 \\ \cdots \\ B\vec{v}_n \end{bmatrix}$$

Where the vector ($\vec{v}$ = payoff_dynamics) components is equal to matrix B. Simply this product is equal to $B\vec{v}$.

*Failure dynamics*: A failure potential can originate at each node with probability $p_n \in [0,1]$, and it also propagates along each link with probability $p_l \in [0,1]$ in each time step. Failure potential turns into a failure with probability $1 - p_p$, depending on an agent's investment into protection; a possible choice is $[p_p = p_{p,max}/(1 + c_{p,1/2}/(f_p c))]$.

Where the protection ($p_p$) is equal to the applied (saturation) function. $p_{p,max}$ is a designated protection maximum, the $c_{p,1/2}$ denotes an allocated reference point, and the $f_p c$ represents an evolutionary protection level multiplied by the updated capital.

$$p_\text{p} = \vec{u}_i = \begin{bmatrix} \vec{u}_{i1} \\ \vec{u}_{i2} \\ \dots \\ \vec{w}_{in} \end{bmatrix}, \quad f_\text{p} = \vec{v}_i \begin{bmatrix} \vec{v}_{i1} \\ \vec{v}_{i2} \\ \dots \\ \vec{v}_{in} \end{bmatrix}, \quad f_\text{p}c = c \begin{bmatrix} c\vec{v}_{i1} \\ c\vec{v}_{i2} \\ \dots \\ c\vec{v}_{in} \end{bmatrix}$$

The applied output given failure dynamics becomes;

$$B = \begin{bmatrix} b_{11} & b_{12} & \cdots & b_{1n} \\ b_{21} & b_{22} & \cdots & b_{2n} \\ \cdots & \cdots & \cdots & \cdots \\ b_{m1} & b_{m2} & \cdots & b_{mn} \end{bmatrix} \begin{bmatrix} \vec{u}_1 \\ \vec{u}_2 \\ \dots \\ \vec{u}_n \end{bmatrix} = \begin{bmatrix} b_{11}\vec{u}_1 + b_{12}\vec{u}_2 + \cdots b_{1n}\vec{u}_n \\ b_{21}\vec{u}_1 + b_{22}\vec{u}_2 + \cdots b_{2n}\vec{u}_n \\ \cdots \cdots \cdots \cdots \\ b_{m1}\vec{u}_1 + b_{m2}\vec{u}_2 + \cdots b_{mn}\vec{u}_n \end{bmatrix} = B\vec{u} = \begin{bmatrix} B\vec{u}_1 \\ B\vec{u}_2 \\ \dots \\ B\vec{u}_n \end{bmatrix}$$

Where the vector ($\vec{u}$ = failure_dynamics) components is equal to matrix B. Simply this product is equal to $B\vec{u}$. Under this section, a pre-written function (Erdös-Renyi) was used to make a short iterate 1=D array for vectorization ($\vec{u}$) instead of using the adjacency matrix directly. This substitution made the loop shorter. Failure lasts for one-time step and causes the loss of an agent's capital.

*Strategy dynamics*: Each agent chooses its protection level according to the heuristics $f_\text{p} = f_\text{p0} + f_\text{p1}C$, truncated to the interval $(0, 1 - f_\text{m})$.

$$\vec{v} \to \vec{f} \to f(\vec{v}), \quad f(\vec{v}) = \begin{cases} 0 < f(\vec{v}) < 0.9, & f_\text{m} = 0.9 \\ 0 < f(\vec{v}) < 0.1, & f_\text{m} = 0.1 \end{cases}, \quad \vec{v}|_{f_\text{p} = f_\text{p0} + f_\text{p1}C}$$

For initialization of the strategy values, two arrays have been added for vectorization $[(f_\text{p0} = \vec{w}_i), (f_\text{p1}C = C\vec{w}_i)]$.

$$f_\text{p0} = \vec{w} = \begin{bmatrix} \vec{w}_1 \\ \vec{w}_2 \\ \dots \\ \vec{w}_n \end{bmatrix}, \quad f_\text{p1}C = C\vec{w} \begin{bmatrix} C\vec{w}_1 \\ C\vec{w}_2 \\ \dots \\ C\vec{w}_n \end{bmatrix}, \quad C \in [0,1]$$

Where the $\vec{w}_{i1}$ is a vectorization as the designated strategy of ($f_\text{p0}$) and the $\vec{w}_{ii}$ is a vectorization as the designated strategy of ($f_\text{p1}$) multiplied by eigenvector centrality from the Erdös-Renyi graph ($C$) which is a measure of the centrality of the agent's node normalized to the interval (0,1).

$$f_\text{p0} = \vec{w}_i = \begin{bmatrix} \vec{w}_{i1} \\ \vec{w}_{i2} \\ \dots \\ \vec{w}_{in} \end{bmatrix}, \quad f_\text{p1}C = C\vec{w}_i \begin{bmatrix} C\vec{w}_{i1} \\ C\vec{w}_{i2} \\ \dots \\ C\vec{w}_{in} \end{bmatrix}, \quad \vec{w}_i + C\vec{w}_i = \begin{bmatrix} \vec{w}_{i1} + C\vec{w}_{i1} \\ \vec{w}_{i2} + C\vec{w}_{i2} \\ \dots \\ \vec{w}_{in} + C\vec{w}_{in} \end{bmatrix} = f_\text{p} = \begin{bmatrix} \vec{v}_{i1} \\ \vec{v}_{i2} \\ \dots \\ \vec{v}_{in} \end{bmatrix}$$

Eigenvector centrality for node i is ($Ax = \lambda x$), where $A$ is the matrix of the network with eigenvalue $\lambda$. The strategy values $f_\text{p0}$ and $f_\text{p1}$ evolve through social learning and strategy exploration as follows. In each time step, each agent with probability $p_\text{r} \in [0,1]$ randomly chooses another agent as a role model and imitates that agent's strategy values with probability ($p_\text{i} = [1 + e^{-\omega \Delta \pi}]^{-1}$, $\pi_r - \pi_f = \Delta \pi|_{\pi_r = \text{role model}}$). Where $p_\text{i}$ is the probability acceptance of the role model for imitation, $\pi_f$ is a capital of the focal

individual, $\pi_r$ is a capital of the role individual, $e$ denotes the exponential, and $\omega$ is the intensity of the selection ($\omega < 1$ = weak selection, $\omega \to \infty$ = strong selection). The focal individual imitates the strategy of the nearby role individual, comparing its new capital (large $\Delta\pi$ = capital difference large, small $\Delta\pi$ = capital difference small), and then the focal individual chooses to imitate the strategy of the role individual. Regarding this imitation part, a temporary matrix was used to avoid changing and using 1 matrix in the loop.

Finally, in each time step, each agent with probability $p_e \in [0,1]$ randomly chooses one of its two strategy values alters it by a normally distributed increment with mean 0 and standard deviation $\sigma_e$ $[f(x|\mu,\sigma^2) = \frac{1}{\sqrt{2\pi\sigma^2}} exp^{-\frac{(x-\mu)^2}{2\sigma^2}} \Big| x = individual\ capital,\ \mu \in R = mean(location),\ \sigma^2 > 0 = variance(squared\ scale)]$.

## 2. Tables of the model parameters

| | | |
|---|---|---|
| Imitation probability | $p_r$ | $\in (0,1)$ |
| Selection intensity | $s$ | $\in (1,100)$ |
| Exploration probability | $p_e$ | $\in (0,1)$ |
| Normally distributed increment | $\sigma$ | $\in (0,1)$ |

**Table 1.** evolutionary part (four parameters).

| | | |
|---|---|---|
| Number of individuals (nodes) | $n$ | $\in (1,100)$ |
| Connection probability | $p$ | $\in (0,1)$ |
| Maintenance | $f_m$ | $\in (0,1)$ |
| Propagation probability at each node | $p_n$ | $\in (0,1)$ |
| Propagation probability through each link | $p_l$ | $\in (0,1)$ |
| Protection maximum | $p_{p,max}$ | $\in (0,1)$ |
| Reference point | $c_{p,1/2}$ | $\in (0,1)$ |

**Table 2.** Non-evolutionary part (seven parameters).

| | | |
|---|---|---|
| Time periods | $t$ | $1 \sim 10{,}000$ |
| Recovery rate | $r\_t$ | $1 \sim 10$ |
| Recovery time delay | $t\_r$ | $1 \sim 10$ |
| Realization | $tt$ | $1 \sim 100$ |

**Table 3.** Time-dependent part (four parameters).

## 3. Code book (Python)

```python
#==============================================================
# CREATED NETWORK
#==============================================================
n = 100                                     # n = number of nodes
p = 0.9                                     # p = connection probabilty [c ∈ (0,1)]
G = nx.erdos_renyi_graph(n, p)              # create random graph
A = nx.adjacency_matrix(G)                  # resultant adjacency matrix

C = nx.eigenvector_centrality(G)            # calculate eigenvector_centraliity.

#==============================================================
# INITIAL PARAMETERS
#==============================================================
B                   = np.zeros((n,4))       # create matrix to store parameters
failure_potential   = np.zeros(n)           # create another matrix to store influence
protection_potential = np.zeros(n)          # create another matrix to store protection potential
fp                  = np.zeros(n)           # create another matrix to store protection level
capital = 1                                 # initial capital

# for strategies
fp0 = np.random.normal(0.7, 0.01, size=[n]) #strategy of the one(mean, SD, size=n)
fp1 = np.random.normal(0.7, 0.01, size=[n]) #strategy of the other(mean, SD, size =n)

# for protection
Centrality = np.array(list(C.values()))     # copying C's values to make an array from a dictionary
fp = B[:,1] + B[:,2] * Centrality           # for protection level
fm = 0.1                                    # for maintenance

# for imitation dynamics
s = 100                                     # selection intensity
pr = 0.9                                    # imitation probability

# for exploration dynamics
pe = 0.05                                   # exploration probability
mu = 0.0                                    # mean for normally increment
sigma = 0.01                                # standard deviation for normally increment

# for failure dynamics
pn = 0.1                                    # with this, a failure potential can originate at each node
pl = 0.1                                    # with this, a failure potential can propagate along each link

# for saturation function
pmax = 1                                    # for protection maximum
cp   = 1                                    # for reference point (=cp,1/2)

# for recovery rate
rec1 = 1.0                                  # always reset (failure potential)
```

```python
    rec2 = 1.0                                          # never reset  (failure)

# fore recovery time delay
failtime  = 1                                           # determines the number of timesteps
failtimear = np.zeros(n)                                # during which a node is failed
failidx   = []                                          # index of failed nodes

# for time steps and realization
timePeriod = 4000                                       # time period
realization = 1                                         # repeats of the simulation

#===============================================================================
# MODEL'S DYNAMICS
#===============================================================================
for real in range(realization):                         #REALIZATION LOOP

    #-------------------------------------------------------
    # Initial condition
    #-------------------------------------------------------
    B[:,0] = capital                                    # initialized capital for all individual within the loop
    B[:,1] = fp0                                        # strategy of the one for all individual within the loop
    B[:,2] = fp1                                        # strategy of the other (fp1*C) for all individual within the loop
    B[:,3] = 0                                          # 0 -> not fail | 1-> fail: initially without failure

    for t in range(0, timePeriod + 1 ):                 # loop for time steps
        temp = B                                        # a temporarily variable to save strategies

        #-------------------------------------------------------
        for i in range(n):                              # loop for every individual within the t loop

            #-------------------------------------------------
            # imitation
            #==========
            """Each agent with probability pr randomly chooses another agent as a role model and imitates that agent's strategy values with probability pi."""
            R1 = np.random.random()                     # randomly choose a certain(%) only 1 time
            if R1 <= pr:                                # conditional
                ff = i                                  # focal model (each node i)
                while True:                             # it is true
                    rr = np.random.choice(n)            # randomly choose role model
                    if ff != rr:                        # until focal choose a different role model
                        break                           # exit out of the loop
                pi = 1 / (1 + (np.exp(-s*(B[rr,0]-B[ff,0]))))   # calculate (fermi) function
                R2 = np.random.random()                 # randomly choose a certain(%) only 1 time
                if R2 <= pi:                            # conditional
                    temp[ff, 1:3] = B[rr,1:3]           # imitate the role model
            B[ : , 1 : 3] = temp[ : , 1 : 3]            # update strategy values
            #-------------------------------------------------
            # exploration
            #============
            """Each agent with probability pe randoly chooses one of its two strategy values and alters it by a normally distributed increment with mean (0) and SD (sigma)"""
```

```python
        temp        =   B[:,1:3]                                    # a temporarily variable to save strategies
        R3          =   np.random.random(size = [n, 2]) <= (0.5 * pe)    # randomly choose a certain(%) with conditional
        temp[R3]    +=  np.random.normal(mu, sigma, size = [n,2])[R3]    # normally distributed increment
        B[:,1:3 ]   =   temp                                        # update strategy values

        #--------------------------------------------------------
        # capital
        #========
        """Each agent chooses its protection level according to the heuristics fp=fp0+fp1*C truncated to the interval (0,1 -fm), where C is a measure of the centrality of the agent's node normalized to the intervalu (0,1)"""
        fp = B[:,1] + B[:,2] * Centrality                           # for heuristics of the protection level
        fp[fp < 0]       = 0                                        # for truncation of the interval (0)
        fp[fp > (1 - fm)] = 1 – fm                                  # for truncation of the interval (0,1 - fm)
        """Each agent receives one unit of payoff, which is added to its capital c, of which fractions fm and fp are spent on maintenance and protection, respectively, resulting in the updated captial 1 + (1 - fm - fp)*c"""
        B[:,0]= 1 + (1 - fm - fp) * B[:,0]                          # resulting in the updated capital

#=================================================================
        # FAILURE DYNAMICS

#=================================================================
        """A failure potential can originate at each node with (pn) ,and can turn into a falure with (1 - pp). Failure potenital propagates along each link with (pl), by a failed node"""
        R4 = (np.random.random(n) <= pn)                            # randomly choose a certain(%) with conditional
        failure_potential[R4] = 1                                   # conditional if it is
        for i in range(n):                                          # loop for every individual within the loop with conditional
            if B[i,3] > 0:                                          # if node failed
                neighbors = nx.all_neighbors(G, i)                  # apply network property
                for j in neighbors:                                 # find out link of the node (neighbor linked by connection)
                    R5 = np.random.random()                         # randomly choose a certain(%)
                    if R5 <= pl:                                    # if it is (conditional)
                        failure_potential[j] = 1                    # do this (failure potential goes to)

        #---------------------------------------------------------------
        # failure potential in neighbors can turn into failure
        #======================================================
        """A failure potential turns into a failure with probability 1-pp, depending on an agent's investment into protection pp=ppmax/((1+cp,1/2)/(fpc))"""

        B[:, 3]   = 0                                               # node failed
        protection_potential = np.zeros(n)                          # conditional
        index     = (failure_potential   >   0 )                    # return as (true or false) for all individuals
        protection_potential[index]      = (pmax / (1 + cp/(fp[index] * B[index,0])))     # index true goes to
        protection_potential[ np.isnan(protection_potential ) ] = pmax        # conditional
        R6        = ((np.random.random(n) <=  1 - protection_potential) & index)   # randomly choose a certain(%) with conditional
        B[R6, 3]  = 1                                               # individuals with index true, becomes 1
```

```python
        B[R6, 0]  = 0                                          # individuals with index true, becomes 0

        #-------------------------------------------------------
        # reset failure potential and/or failures
        #=======================================
        """A failure lasts for one time step and causes the loss of an agent's capital: reset failure potential and/or failure"""
        R7     = np.random.random(n)                            # randomly choose individuals with a certain(%)
        index  = R7 < rec1                                      # indexing with conditional
        failure_potential[index] = 0                            # individuals with index true, becomes 0 (reset the value)
```